\documentclass[a4paper, amsfonts, amssymb, amsmath, reprint, showkeys, nofootinbib, twoside]{revtex4-1}
\usepackage[english]{babel}
\usepackage[utf8]{inputenc}
\usepackage[colorinlistoftodos, color=green!40, prependcaption]{todonotes}
\usepackage[pdftex, pdftitle={Article}, pdfauthor={Author}]{hyperref} 
\usepackage{subcaption}
\usepackage{ragged2e}
\usepackage{ragged2e}
\usepackage[abs]{overpic}
\usepackage{tocloft}
\usepackage{booktabs}
\usepackage{makecell}
\usepackage{makecell}
\usepackage{graphicx}
\usepackage{xcolor}

\begin{document}
\title{ Modulating hydrodynamic flow by modifying the active patch of a colloid}

\author{Om Vandra$^1$, Suhal Siva Ratan T. N.$^1$, Hemant Giri$^2$, Manish Modani$^2$, Vijay Chikkadi$^1$, Raghunath Chelakkot$^3$, Apratim Chatterji$^1$}
    \email[Correspondence email address: ]{apratim@iiserpune.ac.in \\ raghu@phy.iitb.ac.in}
    \affiliation{1. Dept. of Physics, Indian Institute of Science Education and Research, Dr. Homi Bhabha Road, Pune, India-411008}
     \affiliation{2. Nvidia Graphics Pvt. Ltd, Chinappa Layout, Laxmi Sagar Layout, Mahadevpura, Bengaluru, India 560048}
     \affiliation{3. Department of Physics, Indian Institute of Technology Bombay, Powai, Mumbai, India, 400076.}
     

\date{\today} 

\begin{abstract}
We have developed a simulation model to study the hydrodynamic flow fields around Brownian colloidal particles with an active surface patch. Hydrodynamics is introduced by modeling low-Reynolds-number fluid flows around a colloid using multi-particle collision (MPC) dynamics and allowing momentum exchange between the MPC fluid and the colloid. This approach provides good estimates of both near- and far-field flows around the colloid. The size of the active patch is varied to generate different fluid flow fields around the colloid. In this framework, the fluid in the vicinity of the active patch is driven radially away from (or toward) the surface, and an equal and opposite momentum is imparted to the colloid to ensure momentum conservation. The resulting surface-driven flow generates self-propulsion of the particle, thereby converting an otherwise Brownian colloid into an active Brownian particle. Interestingly, as we systematically vary the surface area of the active patch on the colloid, the nature of the generated flow field changes from that of a pusher to a puller. To model such surface activity-driven flows, we developed a hybrid boundary condition that ensures a no-slip condition while incorporating momentum exchange between the flowing fluid and the colloid surface. This scheme integrates the advantages of bounce-back and stochastic boundary conditions while mitigating their respective limitations. Thus, in future studies, the effective hydrodynamic interactions between an active and a passive colloid, or between two active colloids, can be modulated by adjusting the size of the active patch.

\end{abstract}

\maketitle

\section{Introduction}
Swimming microorganisms are ubiquitous in the living world. A wide variety of micro-organisms employ diverse strategies to move through viscous fluids by undergoing periodic, non-reciprocal deformations of their bodies or appendages~\cite{lauga2009hydrodynamics,goldstein_2024}. For example, {\em Escherichia coli} rotates its helical flagellum for swimming, and spermatozoa utilize the beating of a flexible flagellum attached to the rear of their bodies for propulsion. Protists such as {\em Paramecium} swim by coordinated beating of cilia covering their bodies. In contrast, {\em Chlamydomonas}, a biflagellated green algae, propels itself using flagella located at its anterior end. Each distinct swimming strategy creates its own unique hydrodynamic flow fields around the micro-organism. In some cases, such organisms form multicellular colonies, such as {\em Volvox}, which are believed to enhance swimming efficiency and collective functionality~\cite{goldstein2015green}.

Despite the diversity of swimming strategies exhibited by microorganisms, many share common hydrodynamic features that motivate simplified theoretical models~\cite{lauga2016bacterial,ganguly2026hydrodynamics}. The leading-order far-field flow is described by a force dipole, or stresslet, which can be extensile for pushers and contractile for pullers. Ciliated microorganisms such as {\em Protozoa} and {\em Volvox} are often modeled using the squirmer framework~\cite{ganguly2025hydrodynamics,pak2014generalized}, in which ciliary activity is represented by an axisymmetric tangential slip velocity on the swimmer surface~\cite{lighthill1952squirming,blake1971spherical,pedley2016spherical}. The squirmer model has also been widely employed to describe synthetic microswimmers propelled by self-phoretic mechanisms. Although such simplified descriptions have proved useful in theoretical and computational studies, experimentally observed swimmer flows, particularly in the near field, are often substantially more complex~\cite{drescher2010direct,drescher2011fluid,lushi2014fluid,drescher2009dancing,goldstein_2024}.



Experimentally, active motion is frequently investigated using Janus colloids in controlled environments~\cite{gompper20252025,liebchen2018synthetic,saintillan2014theory}. Chemical reactions on the active surface generate local concentration gradients that induce phoretic flows near the particle. Fluid adjacent to the active patch is therefore transported either away from or toward the surface and exchanged with the surrounding bulk fluid~\cite{xiao2025ionic,brady2021phoretic}. The resulting flow generates self-propulsion, transforming a Brownian colloid into an active Brownian particle.

The ability to reproduce diverse hydrodynamic flow fields associated with biological swimmers is important for the design of artificial micro- and nanobots~\cite{palagi2016structured}. In recent years, several such synthetic systems with specific functionalities have been developed~\cite{ju2025technology,palagi2018bioinspired,wang2026advanced,li2026microrobots}, including biocompatible microrobots capable of operating within the human body~\cite{ma2015enzyme,tang2020enzyme,tang2024bacterial}. One possible strategy for controlling the flow generated around these active colloids is through the incorporation of engineered active patches on their surfaces. Experimental realizations of colloids with multiple active patches have already been demonstrated~\cite{wang2019active,popescu2025hydrodynamic,anjali2018shape}. More recently, programmable reconfiguration of hydrodynamic flow fields via phototactic activation of selected surface regions has also been achieved~\cite{rohde2025programmable}. In addition, the propulsion velocities of such active colloids can be tuned by inducing variations in concentration gradients across their surfaces~\cite{xiao2025ionic}.


It is therefore desirable to develop theoretical models that can generate diverse flow fields by tuning the geometry of active surface patches and phoretic mechanisms~\cite{de2020self,scagliarini2020unravelling}. Such models would facilitate the rational design of microswimmers with prescribed flow characteristics. This study is a step in this direction. Thus we develop numerical approaches in which the flow field emerges naturally from local swimmer--fluid interactions, without imposing an {\it a priori} velocity profile as is done in the squirmer model. Such approaches accurately capture both near- and far-field hydrodynamics.




In numerical implementations, swimmer--fluid interactions can be introduced through boundary conditions~\cite{lauga2006microfluidics,chatterji2005combining} or localized stresses while maintaining overall force and torque balance. The model must also avoid unphysical local heating of the fluid caused by swimmer activity. Here, we introduce a numerical scheme based on multiparticle collision dynamics (MPC) in which swimmer--fluid coupling is implemented entirely through the novel hybrid boundary condition, which we introduce in this work. The swimmer is modeled as a patchy colloidal sphere whose activity originates from an embedded active surface patch. We show that a broad range of hydrodynamic flow profiles can be generated by tuning a single parameter, namely the angular extent of the active patch. Furthermore, the near-field flow structure is well resolved in our simulations.

The remainder of the paper is organized as follows. In Sec.~II, we describe the MPC framework, the active-colloid model, and the hybrid boundary condition used for momentum exchange between the colloid and fluid. We also benchmark the boundary-condition schemes using standard protocols~\cite{chatterji2005combining,lobaskin2004new}. In Sec.~III, we present the resulting hydrodynamic flow fields for different active-patch sizes and quantitatively characterize the corresponding fluid velocities near the colloidal surface. Finally, Sec.~IV summarizes the main conclusions.


\section{Model}
We modeled an active colloid as a spherical particle immersed in a fluid described by multiparticle collision (MPC) dynamics. The fluid–colloid interaction was implemented to satisfy the no-slip boundary condition (BC), whereby the fluid velocity at the surface of the colloid equals the velocity of the colloidal surface. To implement this no-slip boundary, two separate schemes are often used: (a) a conventional bounce-back boundary condition, (b) a stochastic boundary condition. Although both methods aim to reproduce no-slip boundary conditions, they do so with varying degrees of fidelity and possess distinct limitations. Here we introduce a hybrid boundary scheme that combines approaches of schemes (a) and (b), and demonstrate that this method mitigates the shortcomings of each scheme.



The methods section is divided into six subsections. 
In the first, we describe the MPC simulation framework, which will be 
used to model hydrodynamic flow fields around colloids. The second subsection gives a brief
summary of the two schemes to model the no-slip boundary conditions at a surface 
in contact with the fluid, and then we 
introduce the new method that we developed.  We refer to the new scheme as 
the hybrid Boundary condition (hybrid-BC).   
In the third subsection, we compare the performance and limitations of different schemes used to model 
the no-slip boundary conditions by simulating Poiseuille flow of an MPC fluid confined between two parallel walls. This also allows us to assess the effectiveness of the hybrid boundary condition.
In the fourth subsection, we describe how we couple the MPC fluid with the colloidal
particle using hybrid-BC, such that there is momentum exchange between the two. This momentum exchange 
results in Brownian motion of the colloid on one hand, and also creates velocity gradients in the MPC
fluid. A consequence of the momentum imparted to the fluid creates the hydrodynamic flow fields around the colloid.  In the fifth subsection, we characterize the distribution of thermal velocities
of the colloid to examine the effectiveness of the colloid-fluid coupling, where we use hybrid-BC on a
spherical surface. We also verify that the fluid temperature remains correctly maintained near the colloidal surface during momentum exchange with the sphere, even without a thermostat.

The final subsection  details the implementation of the momentum exchange 
between an active patch on the colloidal surface and the MPC fluid,  which  models 
The motion of a self-propelled colloid in a hydrodynamic medium.
Furthermore, we show that the temperature of the fluid in the vicinity of the active colloid 
remains within reasonable limits, despite the local injection of energy into the system at the active patch. The hybrid-BC maintains the fluid temperature without the need for a separate thermostat. We further characterize how we can control the colloid's propulsion velocity in the active direction, thereby setting the Peclet number.

\subsection{Modeling Hydrodynamics using MPC}

In the MPC scheme, the fluid is modeled by $N$ point-like particles of mass $m$ moving in free space. Each of these particles are carriers of mass and momentum and can be considered as a coarse-grained 
fluid element. The system evolves through two alternating steps: a streaming step, during which particles move ballistically, and a collision step, during which their velocities are collectively updated. This step mimics
collisions between particles, and this results in momentum exchange between a subset of particles that are proximal. 

During the streaming step, the position of the MPC particles is updated with time $\Delta t$,
\begin{equation}
\label{eq:streaming_step}
\mathbf{r_i}(t + \Delta t) = \mathbf{r_i}(t)+\mathbf{v_i}(t)\Delta t
\end{equation} where, $\mathbf{v_i}(t)$ and $\mathbf{r_i}(t)$ are the velocity and position of the particle $i$ at time $t$, respectively.  During the collision step, the simulation box is first divided into smaller cells (cubic collision boxes) of size a. Subsequently, the velocities of the particles within each cell are rotated in a stochastic manner (detailed hereafter), leading to momentum exchange among the particles in that cell.  

There are multiple schemes for implementing such momentum exchange. In the present work, we employ a specific method known as Stochastic Rotation Dynamics (SRD). For MPC-SRD, the center of mass velocity $\mathbf{v_{cm}^C }(t) $ of particles within the cell $C$ for time $t$ is first computed. The particle velocities relative to $\mathbf{v_{cm}^C} (t) $ are then rotated by a fixed angle $\alpha$ about a randomly oriented axis generated independently for each cell. 
If $\mathcal{R}$ is the rotation matrix which acts on $ (\mathbf{v_i}(t)-\mathbf{v_{cm}^C}(t))$, the particle velocity $\mathbf v_i$ is updated as, 
\begin{equation}
\label{collision_step-a}
\mathbf{v_i}(t + \Delta t) = \mathbf{v_{cm}^C}(t) + \mathcal{R} \cdot (\mathbf{v_i}(t)-\mathbf{v_{cm}^C}(t) ).
\end{equation}
Note that this  collision step conserves energy and 
linear momentum, but it fails to conserve the angular momentum. This absence of angular momentum conservation 
can lead to simulation artifacts in certain systems~\cite{gotze2007relevance}. Therefore, a suitably improved collision 
algorithm that conserves angular momentum along with energy and linear momentum~\cite{noguchi2008transport}, is preferred, and discussed hereafter. In this modified algorithm for the collision rule, we define the center of mass angular velocity {\em  before} rotation, 
$\mathbf{{\omega}_{cm}^C}(t)$, and {\em after} rotation, 
$\mathbf{{\Omega}_{cm}^C}(t)$ at time $t$ as,

\begin{multline}
    \mathbf{{\omega}_{cm}^C}(t) = m \mathbf{\Pi}^{-1}(t) \sum_{i \in C} ((\mathbf{r_i}(t)-\mathbf{r_{cm}^C}(t)) \\ \times (\mathbf{v_i}(t)-\mathbf{v_{cm}^C}(t))) 
\end{multline} and
\begin{multline}
    \mathbf{{\Omega}_{cm}^C}(t) = m \mathbf{\Pi}^{-1}(t) \sum_{i \in C} ((\mathbf{r_i}(t)-\mathbf{r_{cm}^C}(t)) \\ \times \mathcal{R} \cdot (\mathbf{v_i}(t)-\mathbf{v_{cm}^C}(t)))
\end{multline} Here, $\mathbf{r_{cm}^C}(t) $ and $\mathbf{\Pi}(t) $ are the center of mass vector and moment of inertia tensor of particles in cell $C$ at time $t$, respectively. Using these angular velocities, we can define the collision rule as follows:
\begin{multline}
\label{collision_step+a}
\mathbf{v_i}(t + \Delta t) = \mathbf{v_{cm}^C}(t) +
  \left[\mathbf{\omega_{cm}^C}(t) \times (\mathbf{r_i}(t) - \mathbf{r_{cm}^C}(t))\right] \\
  + \phi^C \left[\mathcal{R} \cdot (\mathbf{v_i}(t) - \mathbf{v_{cm}^C}(t))\right] \\
  - \phi^C \left[ \mathbf{\Omega_{cm}^C}(t) \times (\mathbf{r_i}(t) - \mathbf{r_{cm}^C}(t)) \right]
\end{multline} where, 
\begin{equation}
\phi^C = \sqrt{\frac{\sum_{i \in C} \mathbf{u_i}^2(t) }{\sum_{i \in C}\mathbf{u'_i}^2(t)}}. 
\end{equation} There terms $\mathbf{u_i}(t)$ and $\mathbf{u'_i}(t)$ are defined as follows,

\begin{multline}
\mathbf{u_i}(t) = \mathbf{v_i}(t) - \mathbf{v_{cm}^C}(t) - \\ \left[ \mathbf{\omega_{cm}^C}(t) \times (\mathbf{r_i}(t) - \mathbf{r_{cm}^C}(t)) \right] 
\end{multline} and
\begin{multline}
\mathbf{u'_i}(t) = \mathcal{R} \cdot (\mathbf{v_i}(t) - \mathbf{v_{cm}^C}(t)) - \\ \left[ \mathbf{\Omega_{cm}^C}(t) \times (\mathbf{r_i}(t) - \mathbf{r_{cm}^C}(t)) \right]
\end{multline}

In literature, the collision rule of Eq.(\ref{collision_step-a}) is referred to as \textbf{SRD-a}, and the collision rule of Eq.(\ref{collision_step+a}) is referred to as \textbf{SRD+a}~\cite{noguchi2008transport,gompper2009multi,noguchi2007particle}. 

Neither collision rule is Galilean invariant in its original form. If the mean free path ($\lambda$) of the SRD particle is smaller than the cell size $a$, then the distance traveled between collisions is small. As a result, the same set of particles can repeatedly exchange momentum with each other, leading to the buildup of local correlations in velocities. To mitigate this effect, a random shifting of the entire simulation box is performed before the collision step. 
This is implemented by shifting all particles by a three-dimensional vector with random orientation and magnitude, with a maximum magnitude of $a/2$.

\renewcommand{\arraystretch}{1.4}
\begin{table}[h]
\centering
\resizebox{\columnwidth}{!}{
\begin{tabular}{|l|c|c|}
\hline
Simulation Parameters & In Unit  & Value \\
\hline
\hline
Size of collision Cell ($a$)                                 & $a_0$                         & 1             \\
mass of SRD particle ($m$)                                   & $m_0$                         & 1             \\
Average kinetic energy ($\frac{1}{2}mv^2$)                   & $k_BT_0$                        & 1.5 
            \\
Average number density ($\gamma)$                            & $a_0^{-3}$                    & 10     
        \\
Collision time step ($\Delta t$)                             & $ \sqrt{{m_0}{a_0^2}/{k_BT_0}}$ & 0.1           \\
Rotation angle ($\alpha$)                                    & degree                        & $135^{\circ}$ \\
\hline
\end{tabular}
}
\caption{\justifying List of parameters and their value used in our simulations of the SRD fluid. One can choose the unit $a_0$ to be $10^{-6}~m$ if it is a micron-sized colloid, but one can also choose $a_0=1 ~ nm$ for a nano-meter-sized colloid. One can then appropriately choose the unit of mass $m_0$ depending on the colloid's density and size. One can similarly choose the value of $T_0$ and choose the value of the energy unit $k_B T_0$. The units of other quantities get expressed in units of $a_0$, $m_0$, and $k_BT_0$, considering an uncharged colloidal particle. }
\label{srd_parameter}
\end{table} 

The unit of time $\tau_0 = \sqrt{m_0 a_0^2/k_BT_0}$ is chosen by the choice of the mass $m=1$ of an SRD particle, the unit of energy is set by the choice of the thermal energy $k_B T=1$, and the distance or size of the box is measured in units of the length of the cubic collision box $a$, we set $a=1$. The SRD particles move in straight lines for $\Delta t =0.1 ~\tau_0$ in the ballistic step, before undergoing collision in the collision step. It is possible to analytically calculate transport coefficients such as viscosity~\cite{noguchi2008transport} for SRD fluids. All the simulations in this work are performed using the \textbf{SRD+a} scheme (referred to as SRD hereafter), with parameters given in Table~\ref{srd_parameter}. For our choice of parameter values, the analytical value of viscosity is found to be $\eta \approx 4.6 ~m_0/a_0 \tau_{0}$ and a Schmidt number $Sc \approx 8$. The analytical formula is given in Appendix-\ref{appendix_a}.

\subsection{Modeling of Fluid--Solid boundary condition}
\label{Boundary condition}
The implementation of solid boundaries, either stationary or moving relative to the surrounding fluid, arises in a broad class of fluid-dynamics problems.
In this section, we discuss three methods for coupling boundaries to the SRD fluid. The first two schemes, namely, bounce-back and stochastic BC, are well established in the literature~\cite{inoue2002development,malevanets1999mesoscopic,whitmer2010fluid}. We propose a third approach as an alternative to the previous two methods, combining the advantages of both. 

The fluid in direct contact with a stationary surface has zero velocity, a condition known as the no-slip boundary condition. For a moving surface, this condition implies that the tangential velocity of the fluid layer in contact with the surface equals that of the surface. Moreover, the normal component of the fluid velocity at the confining surface must vanish, ensuring that SRD particles neither penetrate the surface nor emerge from it.  We numerically implement this as follows~\cite{inoue2002development,hecht2005simulation}.

At the end of the streaming step, if particle $i$ crosses the wall, it is repositioned to the nearest point on the surface relative to its previous position.
This point on the wall is taken as the approximate point of impact, denoted by $\mathbf{r^*_i}$. A new velocity, $\mathbf{v'_i}$, is then assigned to this particle using one of three methods: bounce-back, stochastic, or hybrid, each of which is described in detail below. The SRD particle is subsequently moved with the updated velocity $\mathbf{v'_i}$ for time $\Delta t/2$.

Note that the method discussed above is an approximation. A more precise approach would require calculating the exact time of impact rather than approximating it as $\Delta t/2$. This also includes determining the exact point of contact on the wall surface. For simple geometries, such as flat surfaces, both the impact time and location can be computed exactly. However, since our objective is to model collisions between SRD particles and a moving colloidal surface, the chosen approximation offers a significant advantage in reducing computational cost while maintaining sufficient accuracy.

Below, we list all three Boundary conditions (BC) in detail:



\begin{enumerate}

    \item \textbf{Bounce-back boundary condition:} For this condition, a new velocity is generated by reversing the old velocity:

    \begin{equation}
        \mathbf{v'_i} = -\mathbf{v_i} + 2\mathbf{u_s}
    \end{equation}

    Here, $\mathbf{u_s}$ is the velocity of the surface at the point of collision ($\mathbf{r^*_i}$). This boundary condition provides a fairly accurate implementation of the no-slip condition compared to other available methods, but a simple calculation suggests that, for $\mathbf{u_s} \neq 0$, the kinetic energy of the SRD particles increases over the iterations. Hence, for such cases, one needs to use a thermostat with this boundary condition. 

    \item \textbf{Stochastic boundary condition:} For this condition, a new velocity is generated randomly from the Gaussian distribution. Let $\mathbf{\hat{n}}$ be the normal to the surface at the point of impact $\mathbf{r^*_i}$ of the SRD particle $i$ with the surface. Two random unit vectors, $\mathbf{\hat{t}_1}$ and $\mathbf{\hat{t}_2}$, are generated such that they are mutually orthogonal and each is orthogonal to $\mathbf{\hat{n}}$. Then,

    \begin{equation}
        \mathbf{v'_i} = v_1\mathbf{\hat{t}_1} + v_2\mathbf{\hat{t}_2} + \sqrt{v_3^2 + v_4^2} \mathbf{\hat{n}} + \mathbf{u_s}
        \label{eq:stochastic_update}
    \end{equation}

    Here, $v_1,v_2,v_3$ and $v_4$ are four random numbers drawn from the Gaussian random distribution with zero mean and variance $\sqrt{k_BT/m}$. Refer to Appendix-\ref{appendix_b} for a discussion on why four random numbers are needed in Eq.(\ref{eq:stochastic_update}). This coupling scheme more accurately mimics collisions between a molecular fluid and a surface. Due to the finite roughness of surfaces at the length scale of fluid molecules, fluid particles can undergo multiple collisions with the surface. As a result, the outgoing velocities become effectively uncorrelated with the incoming ones. 
     
     Since the outgoing velocities are sampled from the Gaussian distribution, an additional thermostat is not needed when this method is used to model the collision of a fluid particle with a surface. However, this scheme does not enforce the no-slip boundary condition accurately.~\cite{whitmer2010fluid,bolintineanu2012no}

    \item \textbf{Hybrid boundary condition:} In this method, we combine aspects of the bounce back and the stochastic boundary condition. The bounce-back rule is applied only to the tangent component of the velocities by reversing it during the collision step, while the normal component of the velocity is generated according to the stochastic coupling scheme. Thus, the updated velocity $\mathbf{v'_i}$ is computed as,

    \begin{equation}
        \mathbf{v'_i} = (-\mathbf{v_i} +\mathbf{u_s} + v_n \mathbf{\hat{n}}) + \mathbf{u_s} + \sqrt{v_1^2 + v_2^2} \mathbf{\hat{n}}
    \end{equation}

    Here, $v_n = (\mathbf{v_i}-\mathbf{u_s})\cdot \mathbf{\hat{n}}$ and $v_1$ and $v_2$ are generated randomly from the Gaussian distribution with zero mean and variance $\sqrt{k_BT/m}$.
    
    Since this method combines features of both the Bounce-back and the Stochastic coupling scheme, it offers the advantages of both. When using this method, a thermostat is not needed because the normal component is generated stochastically. We can also simulate a no-slip boundary condition more accurately, as the tangential component is generated by reversing the tangential component of the old velocity.
    
\end{enumerate}

\subsection{Characterizing the boundary conditions in Poiseuille flow}
To benchmark the performance of the various boundary conditions, we simulated Poiseuille flow of the SRD fluid between two infinite parallel walls. Poiseuille flow is particularly suitable for this purpose because it is straightforward to model and admits an exact analytical solution for no-slip boundary conditions at the walls. This enables a direct comparison between the simulated velocity profiles and the theoretical prediction, thereby providing a quantitative assessment of the effectiveness of the boundary conditions.




\begin{figure}[]
    \centering

    \begin{subfigure}{0.85\linewidth}
        \centering
        \begin{overpic}[width=\linewidth]{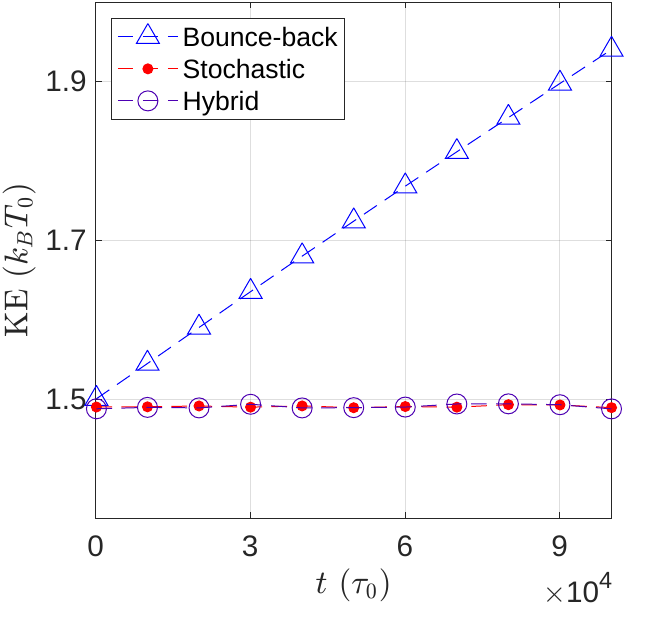}
        \end{overpic}
    \end{subfigure}

    \caption{\justifying Mean kinetic energy (KE) of the SRD fluid for three boundary conditions: bounce-back, stochastic, and hybrid. The KE is shown as a function of time during Poiseuille flow without a thermostat. The flow is along $\mathbf{\hat{x}}$, with confining walls at $z/a = 1$ and $z/a = 49$ in a $50a \times 50a \times 48a$ simulation box with periodic boundary conditions in the $\mathbf{\hat{x}}$ and $\mathbf{\hat{y}}$ directions. For the bounce-back condition, the kinetic energy increases linearly with time, whereas for the stochastic and hybrid boundary conditions, it remains constant despite the absence of a thermostat. }
    \label{fig:poiseuill_temp}
\end{figure}

\begin{figure*}[t]
    \centering
    \begin{overpic}[width=0.32\linewidth]{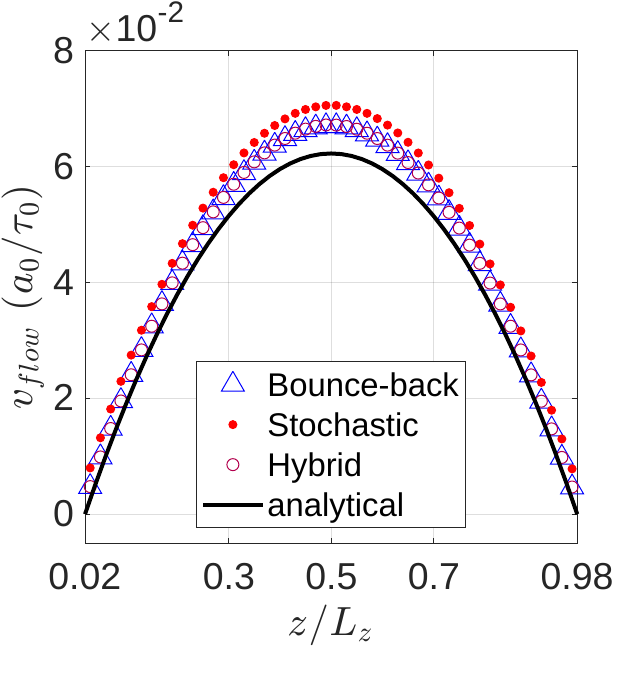}
        \put(125,145){\textbf{(a)}}
    \end{overpic}
    \hfill
    \begin{overpic}[width=0.32\linewidth]{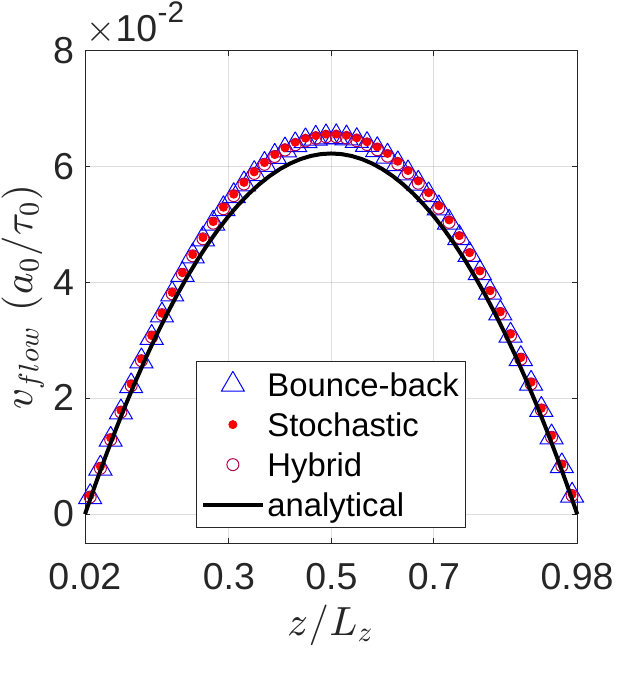}
        \put(125,145){\textbf{(b)}}
    \end{overpic}
    \hfill
    \begin{overpic}[width=0.32\linewidth]{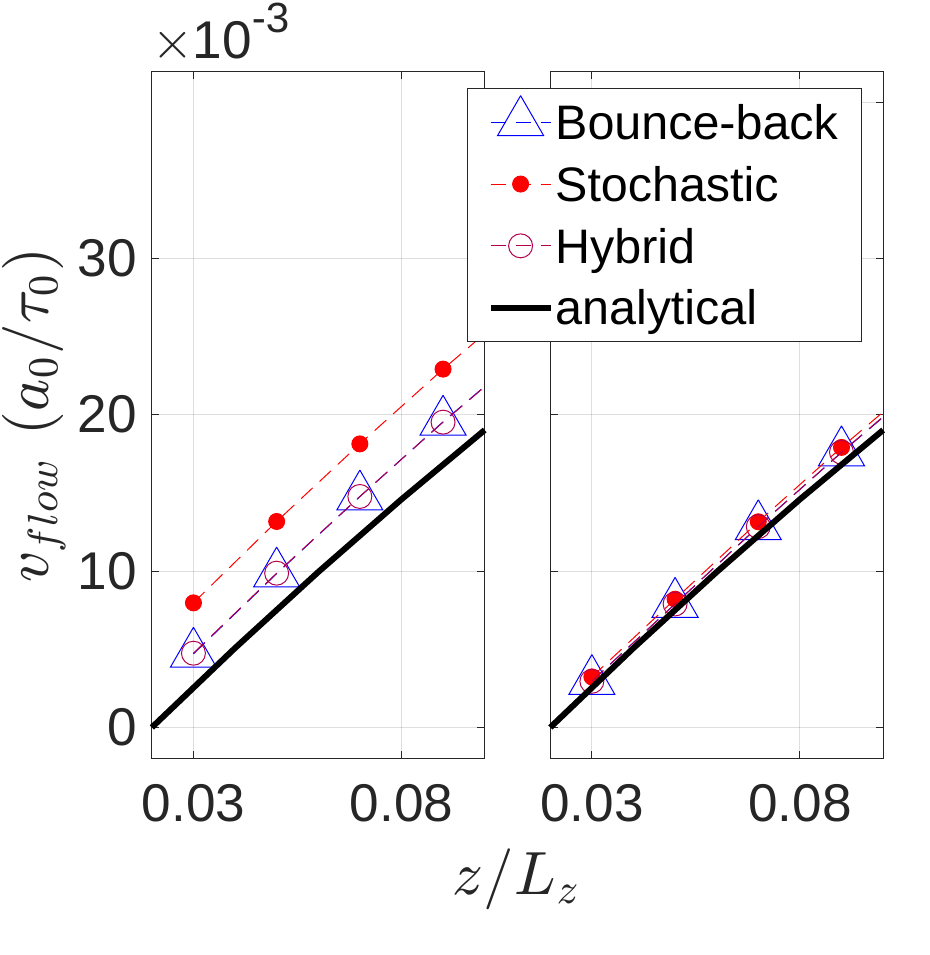}
        \put(65,45){\textbf{(c)}}
        \put(135,45){\textbf{(d)}}
    \end{overpic}
    \caption{\justifying  
    (a) Poiseuille flow velocity profiles $v_{\mathrm{flow}}$ for bounce-back, stochastic, and hybrid boundary conditions without virtual particles. A thermostat is applied only for the bounce-back case. (b) Corresponding profiles with virtual particles at the walls; here, a thermostat is applied for all three boundary conditions. The analytical no-slip profile is shown for reference. Panels (c) and (d) present zoomed-in views of (a) and (b), respectively, near the walls, demonstrating that the inclusion of virtual particles yields an almost no-slip velocity at the boundaries.
   }
    \label{fig:poiseuill}
\end{figure*}

To model this flow, two parallel flat walls are positioned at $ z = a $ and $ z = L_z - a $ within a simulation box of dimensions $ L_x \times L_y \times L_z $. The walls interact with the SRD fluid through one of the three methods described in Sec.~\ref{Boundary condition}. Periodic boundary conditions are applied along the 
$x$ and $ y$ directions, and we introduce an active force on each fluid element in the $\mathbf{\hat{x}}$-direction ~\cite{bolintineanu2012no}. Thus, the position and velocity coordinates are updated as
\begin{gather}
    \mathbf{r_i}(t+\Delta t) = \mathbf{r_i}(t)+\mathbf{v_i}(t)\Delta t + \frac{1}{2} g {\Delta t}^2\mathbf{\hat x} \\
    \mathbf{v_i}(t+\Delta t) = \mathbf{v_i}(t)+g\Delta t
\end{gather}
Here, $g$ is the force applied on each SRD-particle of unit mass. We use $g=0.0001~k_BT_0/a_0m_0$ and a simulation box of $50a \times 50a \times 50a$.

During the implementation of random shifts, the collision cells intersecting the walls 
({\em wall-cut cells}) typically contain fewer SRD particles than those in the bulk. Because of this, the effective viscosity of SRD-fluid will reduce in the {\em wall-cut} cells, causing a finite slip at the wall. To eliminate slip velocity at the boundaries, it is necessary to introduce virtual particles~\cite{lamura2001multi}.  These particles are analogous to SRD particles but are artificially inserted into collision cells that intersect a wall. For completeness, we investigate the Poiseuille flow system both with and without virtual particles.

For the case of Poiseuille flow simulations, we introduce $k_p$ virtual particles 
the $p-$th wall-cut cell in the regions just beyond the bounding walls, specifically in the regions $0 < z < a$ and $L_z - a < z < L_z$. These particles are placed uniformly within these regions to maintain a number density of $\gamma$ for each collision box adjacent to the wall as described in the reference~\cite{zottl2018simulating}. Their velocities are assigned from a Gaussian distribution with a mean of zero and a standard deviation of $\sqrt{k_B T / m}$ for each component of velocity. After initialization, these virtual particles interact with the MPCD particles using the standard SRD+a collision scheme as described by Eq.(\ref{collision_step+a}). 


We first examine the effect of the three BCs on the mean kinetic energy to characterize the thermal energy of the SRD particles. In calculating the mean kinetic energy, we consider only fluctuations about the local mean velocity of the particles. Note that the mean velocity in each layer depends on the distance from the walls, i.e., the fluid velocity is $z$-dependent.  Fig.~\ref{fig:poiseuill_temp} shows the time evolution of the mean kinetic energy for all three BCs without virtual particles and without the use of a thermostat. As discussed in Sec.~\ref{Boundary condition}, the fluid temperature increases steadily for the bounce-back boundary condition, as particles continuously gain kinetic energy from the driving force. In contrast, the mean kinetic energy remains constant for systems with stochastic or hybrid boundary conditions, since the walls effectively act as a thermostat. 

Next, we calculate the fluid flow profile to compare the effective slip velocity at the wall for each of the three BCs mentioned above. Fig.~\ref{fig:poiseuill}(a) shows the fluid velocity profile, $v_{\mathrm{flow}}(z)$, obtained without the inclusion of virtual particles. For this purpose, the fluid velocity was averaged over $5 \times 10^6$ iterations after the flow had reached steady state following an initial $5 \times 10^6$ iterations. We have also studied all three BCs with virtual particles, which is shown in Fig.~\ref{fig:poiseuill}(b).  For this figure, we additionally applied a thermostat in which the particle velocities were rescaled so that the mean kinetic energy in each cell, computed after subtracting the local mean velocity, to maintain the right temperature. Fig.~\ref{fig:poiseuill}(c) presents a zoomed-in view of the velocity profile shown in Fig.~\ref{fig:poiseuill}(a), in the vicinity of the wall. As discussed in Sec.~\ref{Boundary condition}, the slip velocity is higher for the stochastic boundary condition compared to the Bounce-back boundary conditions or hybrid-BC.  One can observe in Fig.~\ref{fig:poiseuill}(b) (for zoomed-in profile near the vicinity of the wall, refer to Fig.~\ref{fig:poiseuill}(d)), that Poiseuille flow simulations with inclusion of virtual particles show a significant reduction in the slip. 

\subsection{Colloidal hydrodynamics: hybrid boundary condition (hybrid-BC) on a spherical colloid}
\label{colloidal suspension}
In the previous section, we showed that, although an external thermostat is not required for either the hybrid or stochastic BC, the hybrid-BC leads to lower wall slip in the absence of wall particles. Since implementing wall particles for spherical colloids can be computationally expensive, we therefore use the hybrid-BC for the remainder of our study.

In our system, colloidal particles (both active and passive) were modeled as motile spheres with radius $R_{col}$ and mass $M_{col}$. The coordinates of the colloid are updated by  Molecular dynamics (MD) using the velocity Verlet algorithm~\cite{allen2017computer}. Other than the conservative forces between colloidal particles usually considered in MD, we also considered forces acting due to momentum exchange with the SRD fluid in the equations of motion. In this paper, we provide results from simulations of a single colloid. However, when we will consider multiple colloids in the future, conservative forces due to interactions between colloids will also contribute.



We describe below the steps used in the simulation procedure to incorporate momentum exchange between the spherical colloid and the fluid using hybrid-BC, and thereby describe colloidal dynamics. The particles that are within a distance $a$ from the colloidal surface are considered proximal to the colloid. In hybrid-BC, we update the positions and velocities of both the colloid and the SRD  particles proximal to the colloid as described now. 

If the positions of all the SRD particles in the box are advanced using the streaming step with $\Delta t$,  then a few SRD particles may cross the colloidal surface. Hence, we reposition the particles that have entered the colloid to a point just outside the colloid. To reduce penetration into the colloid, we stream the SRD particles proximal to the colloid with a reduced time step $\Delta t_{md} = \Delta t/n_{md}$, where $n_{md}$ is an integer. We update the colloidal positions using the same reduced time step  $\Delta t_{md}$.

Thereafter, we modify the velocities of the SRD particles that entered the colloid, thereby altering their momentum. To ensure momentum conservation during these interactions, the negative of this momentum change is transferred to the colloidal particle. Finally, the colloid’s velocity is updated using the velocity-Verlet algorithm.

The above hybrid-BC procedure is performed  $n_{md}$ times for the SRD particles proximal to the colloid, before each SRD collision step for all the particles is performed   Eq.~\ref{collision_step+a}.  We also ensure that all the Verlet steps are performed with a time step $ \Delta t_{md} $, and only thereafter is the  SRD-collision step performed at time step $ \Delta t $. For this project, we use $n_{md} = 10$, $M_{col} = 650~m_0$ and $R_{col} = 2.5~a_0$.  The above procedure is described in detail below in a step-by-step manner. The detailed step-by-step procedure of the hybrid-BC is described below.

\

\noindent {\textbf{Step 1:}} We  update the position of the colloid using:
\begin{equation}
    \mathbf{R}(t+ \Delta t_{md}) = \mathbf{R}(t) + \mathbf{V}(t) \Delta t_{{md}} 
    + \frac{\Delta t_{{md}}^2}{2 M_{{col}}} \mathbf{f}(t).
    \label{eq:verlet_position}
\end{equation} Here, $ \mathbf{R}(t) $ and $ \mathbf{V}(t) $ are the position and the velocity of the center of the colloid at time $t$, respectively. Here, $\mathbf{f}(t) $ denotes the external or conservative forces acting on the colloid. These may arise from factors such as gravity or from DLVO interactions with other colloids present in the simulation. 

\ 

\noindent {\textbf{Step 2:}} We then perform the streaming step of the SRD-particles as described by Eq.(\ref{eq:streaming_step}), but with time step $\Delta t_{md}$. i.e.,

\begin{equation}
\label{eq:streaming_step_md}
\mathbf{r_i}(t + \Delta t_{md}) = \mathbf{r_i}(t)+\mathbf{v_i}(t)\Delta t_{md}
\end{equation}

\ 

\noindent {\textbf{Step 3:}} Following the streaming step, we implement the hybrid-BC, and give the SRD particles new velocities at the surface of the colloid, as described in section~\ref{Boundary condition}. It is germane to note that the velocity at the colloid's surface, $\mathbf{u_s}(t)$, also has a rotational component in addition to the center-of-mass velocity. Thus, the net velocity $\mathbf{u_s}$ at the impact point $\mathbf{r_i^{*}}$ of the $i^{th}$ SRD particle on the surface of the colloid is calculated as,
\begin{equation}
    \mathbf{u_s}(t) = \mathbf{V}(t) + (\mathbf{\Omega}(t) \times (\mathbf{r_i^*}(t)-\mathbf{R}(t))),
\end{equation}
where $\mathbf{\Omega}(t)$ is the angular velocity of the colloid. 

\ 

\noindent {\textbf{Step 4:}} After updating the velocity of SRD particles according to hybrid-BC,  we impart equal and opposite linear and angular momentum of all the interacting SRD particles to the colloid, thereby allowing momentum exchange between the colloid and the fluid, and thereby couple the colloidal motion and the fluid velocity field. If $\mathbf{V'}(t)$ and $\mathbf{\Omega '}(t)$ are the colloidal velocity and angular velocity after the momentum exchange step. Then,

\begin{gather}
    \mathbf{V'}(t) = \mathbf{V}(t) - \frac{1}{M_{col}} \sum_i m(\mathbf{v'_i}(t) - \mathbf{v_i}(t))   \\ 
    \mathbf{\Omega '}(t) = \mathbf{\Omega}(t) - \frac{1}{I_{col}} \sum_i m(\mathbf{r^*_i}(t) - \mathbf{R}(t)) \times (\mathbf{v'_i}(t) - \mathbf{v_i}(t))
    \label{eq:momentum_transfer}
\end{gather} Here, $I_{col} = (2/5)M_{col} R_{col}^2$ is the moment of inertia of the colloid.  
Note that in the above equations, the sum over $i$ is performed over only the SRD particles that have crossed the surface of the colloid in the single iteration.

\ 

\noindent {\textbf{Step 5:}} Finally, we update the translational velocity of the colloid as in the velocity Verlet integration scheme. For this, calculate the updated force on the colloid $ \mathbf{f}(t+ \Delta t_{md}) $ and use it to update the velocity,
\begin{multline}
    \mathbf{V}(t+\Delta t_{md}) = \mathbf{V}'(t) + \\ \frac{\Delta t_{{md}}}{2 M_{{col}}} \left[ \mathbf{f}(t+ \Delta t_{md}) + \mathbf{f}(t) \right]
    \label{eq:verlet_velocity}
\end{multline}
Note that $\mathbf{f}$ does not include the forces due to momentum exchange 
with the SRD fluid. 


\begin{figure}[t]
    \centering

    \begin{subfigure}{0.49\linewidth}
        \centering
        \begin{overpic}[width=\linewidth]{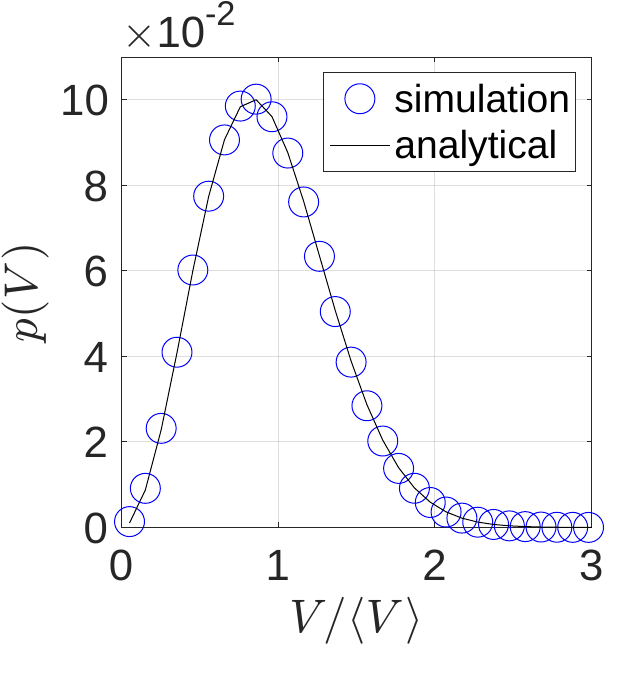}
            \put(25,105){\textbf{(a)}}
        \end{overpic}
        \phantomcaption
        \label{fig:mbp_vel}
    \end{subfigure}
    \hfill
    \begin{subfigure}{0.49\linewidth}
        \centering
        \begin{overpic}[width=\linewidth]{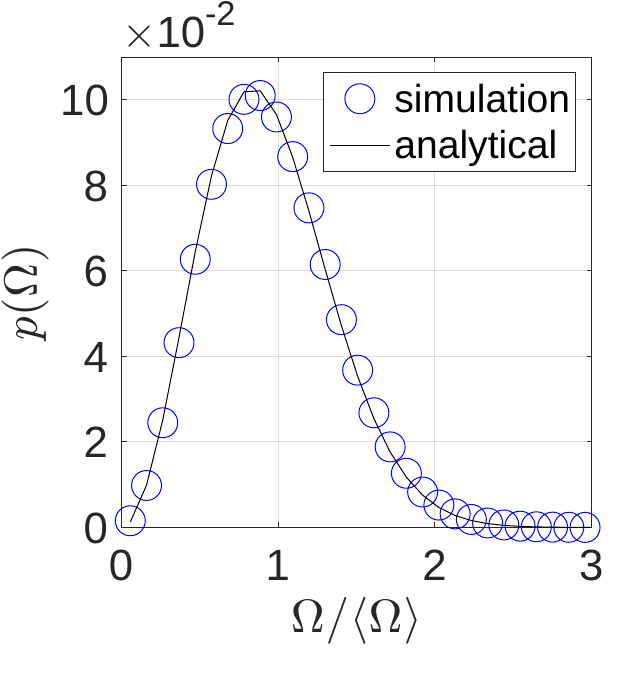}
            \put(25,105){\textbf{(b)}}
        \end{overpic}
        \phantomcaption
        \label{fig:mbp_ang}
    \end{subfigure}

    \caption{\justifying(a) Probability distributions $p(V)$ and $p(\Omega)$ of the translational and rotational speeds of a passive colloid coupled to an MPCD fluid. The speeds are normalized as $V/\langle V \rangle$ and $\Omega/\langle \Omega \rangle$, where $\langle V \rangle = \sqrt{3 k_B T / M_{\mathrm{col}}}$ and $\langle \Omega \rangle = \sqrt{3 k_B T / I_{\mathrm{col}}}$ with $k_B T = 1~k_BT_0$. The solid black lines denote the corresponding Maxwell--Boltzmann distributions.}
    \label{fig:mbp}
\end{figure}

\noindent {\textbf{Step 6:}} Active Brownian particles (such as Janus colloids) are characterized by an orientation vector, which is the direction of active motion of the particles. Since this direction will change over time due to rotational Brownian motion, we need to update their orientation along with their position. Therefore, the orientation vector $\mathbf{\hat{e}}$ is updated as,

\begin{equation}
    \mathbf{\hat{e}}(t + \Delta t_{md}) = \mathbf{\hat{e}}(t) + \left[ \mathbf{\Omega'}(t) \times \mathbf{\hat{e}}(t) \right] \Delta t_{md}.
\end{equation}
Note that the angular velocity has already been updated in Eq.(\ref{eq:momentum_transfer}),  and 
thereby we do not need an explicit equation to update $\mathbf{\Omega}(t)$, as is required in dry systems. After repeating the above six steps $n_{md}$ times, the collision step as described by Eq.(\ref{collision_step+a}) is performed once. The sequence of the six steps is then 
repeated again. 

\begin{figure}[t!]
    \centering

    \begin{subfigure}{0.85\linewidth}
        \centering
        \begin{overpic}[width=\linewidth]{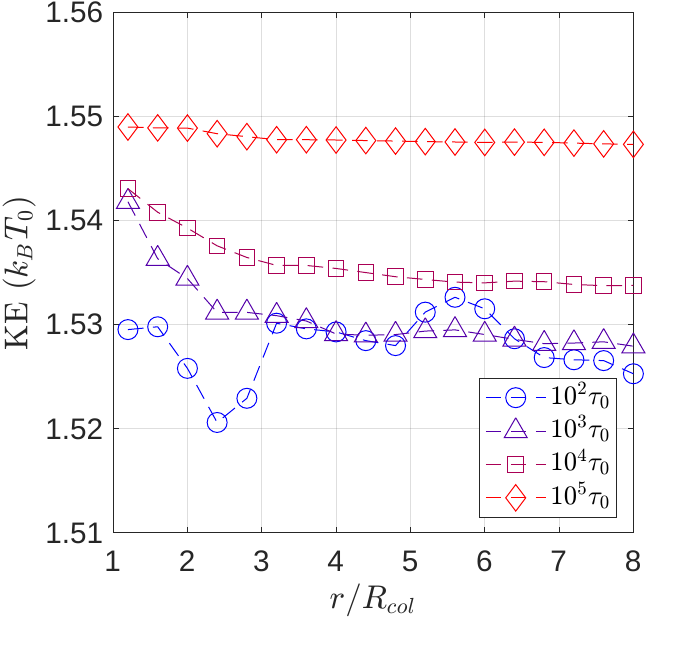}
            \put(55,180){\textbf{(a)}}
        \end{overpic}
    \end{subfigure}

    \caption{\justifying We show the mean Kinetic energy (KE) of the SRD-fluid in the vicinity of a passive colloid.  The  KE was computed within concentric spherical shells of thickness $a$ centered on the colloid and averaged over different time intervals (see legend).  KE is plotted as a function of radial distance $r$ from the center of the colloid.}
    \label{fig:t_shell_pasv}
\end{figure}

We clarify that virtual particles were not introduced for the implementation of the hybrid-BC
on the spherical colloid in order to reduce computational cost. While the use of virtual 
particles is relatively computationally inexpensive in planar geometries, 
it is not so for spherical geometries.

\subsection{ Colloid suspended in MPC fluid: Thermal behavior}
To study the effects of the Hybrid boundary condition on a passive spherical colloid suspended in an SRD fluid, we study the suspension of a single passive colloid in a simulation box of dimension $50a \times 50a \times 50a $ with the periodic boundary conditions.

We provide the speed distribution of translational and angular velocities of the colloid. 
This is shown in Fig.~\ref{fig:mbp_vel} and~\ref{fig:mbp_ang}, respectively. In the figure, we observe that the distribution matches the theoretical Maxwell-Boltzmann distribution exactly at $k_BT = 1 ~k_BT_0$, indicating that the colloid's temperature is maintained.

\begin{figure}[t!] 
    \centering
    \begin{subfigure}{0.45\linewidth}
        \includegraphics[width=\linewidth]{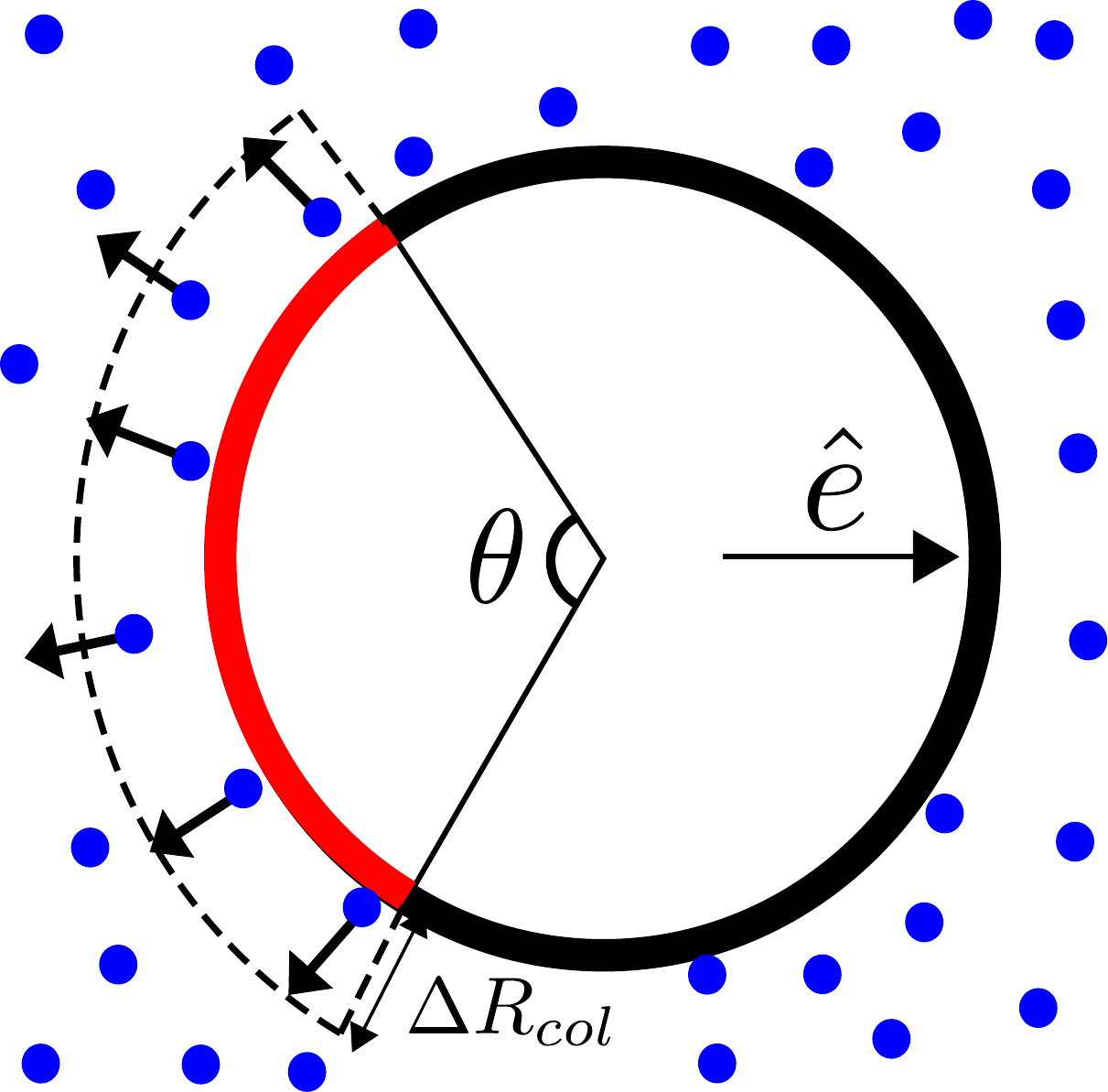}
        \subcaption{Outward pumping}
        \label{fig:schematic_activity_out}
    \end{subfigure}
    \hfill
    \begin{subfigure}{0.45\linewidth}
        \includegraphics[width=\linewidth]{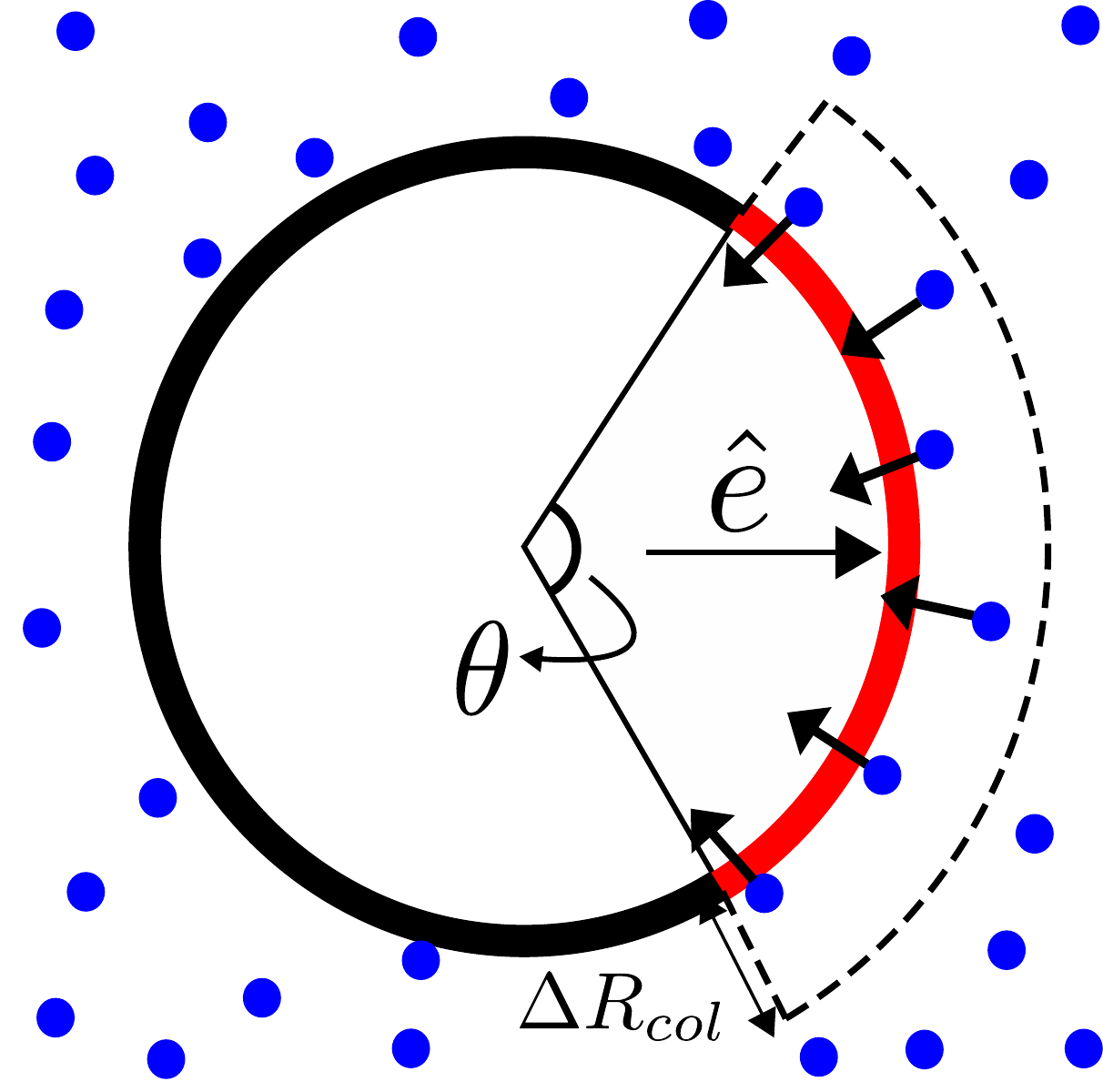}
        \subcaption{Inward pumping}
        \label{fig:schematic_activity_in}
    \end{subfigure}
    \caption{\justifying Schematic of cross section of the active colloid model in an SRD fluid (blue circles). The colloidal surface is partitioned into an active region (red arc) and an inert region (black arc). Activity is introduced by imparting radial momentum, either outward or inward (black arrows), to SRD particles within a distance $\Delta R_{\mathrm{col}}$ from the active surface, corresponding to outward and inward pumping, respectively. For the purpose of this project, we use $\Delta R_{\mathrm{col}} = 0.4a$ for both outward and inward pumping. The active region is characterized by an angle $\theta$ at the colloid center, which is twice the polar angle with respect to the axis $\mathbf{\hat{e}}$.}
    \label{fig:schematic_activity}
\end{figure}

We also establish that the hybrid-BC governing collisions between fluid particles and the colloid properly maintain the fluid temperature in the vicinity of the spherical surface. To characterize the temperature of the SRD fluid as they interact with the surface of the sphere, we plot the time-averaged kinetic energy (KE) of SRD particles in concentric shells of thickness $a$. We chose the center of the passive colloid to be the center of the concentric shells. In Fig.~\ref{fig:t_shell_pasv}, the time-averaged kinetic energy of MPCD particles for shells of different radius averaged over time windows of  $ 10^2~ \tau_0, 10^3 ~\tau_0, 10^4 ~\tau_0, 10^5 ~\tau_0$ is shown.  This plot shows that the average kinetic energy fluctuates around values very close to $1.5~ k_B T_0$ throughout the simulation for the passive colloid.


\subsection{Introducing activity}

To model an active colloid, we partition the surface of a spherical particle into two regions: an active patch with which we hope to mimic the effects of ionic flow due to reaction on the surface, and an inert region. The active region is characterized by the parameter $\theta$, which is twice the polar angle measured with respect to the orientation vector $\mathbf{\hat{e}}$. Thus, $\theta$ denotes the angular spread and hence, the area of the active patch on the surface of the colloid, $2\pi R^2 \int_0^{\theta/2} \sin \theta^\prime d\theta^\prime$. Fig.~\ref{fig:schematic_activity} schematically illustrates this, with the active region shown in red and the inert region in white. Thus, the parameter $\theta$ can be tuned to change the colloid's active surface area.

To implement the activity, we first identify all the SRD-particles within a spherical shell of thickness $ \Delta R_{col} $ surrounding the active surface. In Fig.~\ref{fig:schematic_activity}, these are shown as the blue particles encompassed 
within the dashed outline. At each iteration, all MPCD particles located within this shell are assigned an additional velocity $ v_a $, directed either radially outward (refer to Fig.~\ref{fig:schematic_activity_out}) or inward (see Fig.~\ref{fig:schematic_activity_in}), with respect to the center of the colloid. We refer to these two methods of introducing activity as outward-pumping and inward-pumping, respectively. Notice that the position of the active surface relative to the direction of the orientation vector ($\mathbf{\hat{e}}$) of the active colloid is different for outward and inward pumping, as shown in Fig.~\ref{fig:schematic_activity}.

\begin{figure}[t]
    \centering
    \includegraphics[width=0.85\linewidth]{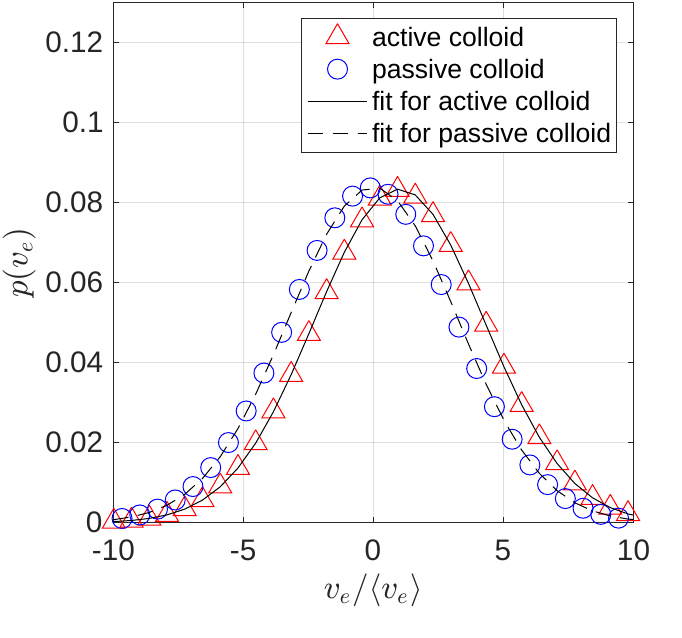}
    \caption{\justifying Probability distribution of the velocity component $v_e$ along the active direction for an outward-pumping colloid with $\theta = 180^\circ$. For comparison, the distribution for a passive colloid, which follows the Maxwell--Boltzmann form, is also shown. The data are well described by Gaussian fits with variance $k_B T = 1~k_BT_0$. The active colloid exhibits a nonzero mean $\langle v_e \rangle$.} 
    \label{fig:mba}
\end{figure}

The total momentum imparted to the MPCD particles within the active shell is computed at every iteration, and an equal and opposite momentum is added to the momentum of the colloid to enforce momentum conservation. Owing to the choice of geometry of the active region, the negative of the total momentum transferred to the MPCD particles is, on average,  aligned with the orientation vector $ \mathbf{\hat{e}}$. This also means that $\mathbf{\hat{e}}$ is also the direction of active motion. This can be demonstrated by plotting the probability distribution of colloidal velocity. 

\begingroup
\renewcommand{\arraystretch}{1.4}
\setlength{\tabcolsep}{10pt} 
\begin{table}[]
\centering
\normalsize 
\begin{tabular}{|c|c|c|c|c|}
\hline
                  & \multicolumn{2}{c|}{Outward Pumping} & \multicolumn{2}{c|}{Inward Pumping} \\
\hline
$\theta$ & $v_a $ & $\langle \mathbf{V} \cdot \mathbf{\hat{e}} \rangle$ & $v_a$ & $\langle \mathbf{V} \cdot \mathbf{\hat{e}} \rangle$ \\
\hline
\hline
$300^{\circ}$ & 0.032  & 0.0119 & 0.032 & 0.0125 \\
$270^{\circ}$ & 0.016  & 0.0124 & 0.016 & 0.0125 \\
$240^{\circ}$ & 0.011  & 0.0123 & 0.011 & 0.0122 \\
$180^{\circ}$ & 0.008  & 0.0123 & 0.008 & 0.0125 \\
$120^{\circ}$ & 0.011  & 0.0123 & 0.011 & 0.0123 \\
$90^{\circ}$  & 0.016  & 0.0121 & 0.016 & 0.0122 \\
$60^{\circ}$  & 0.032  & 0.0116 & 0.032 & 0.0113 \\
\hline
\end{tabular}
\caption{\justifying Values of $v_a$ for different $\theta$ to maintain the value of $\langle v_e \rangle = \langle \mathbf{V} \cdot \mathbf{\hat{e}} \rangle$ close 
to $0.012 ~ a_0/\tau_0$.}
\label{v_a values}
\end{table}
\endgroup 

In Fig.~\ref{fig:mba}, we show the probability distribution of the component of the colloidal velocity along the orientation vector, $v_e= \mathbf{V} \cdot \mathbf{\hat{e}}$. For the data of the velocity distribution shown in Fig.~\ref{fig:mba}, we have considered an active colloid with outward pumping with $\theta = 180^\circ$. We have also plotted the equilibrium probability distribution of the velocity for the passive colloid along one direction in Fig.~\ref{fig:mba} as a reference. We observe that for the choice of $v_a$ values that we have used,  the velocity distributions follow a normal distribution with standard deviation $\sqrt{k_BT/M_{col}}$, with $k_BT = 1 ~k_BT_0$ for both the active and the passive colloid. As expected, the mean for the passive colloid is zero. But, for active colloid, we observe that $v_e$ follows a normal distribution with mean $ \langle v_e \rangle = 0.0123 ~ a_0/\tau_{0}$  for a choice of  $v_a = 0.008 ~a_0/\tau_0$.

The momentum transferred to the colloid will depend on the values of $v_a$, $\theta$, and $ \Delta R_{col}$.  That in turn determines the mean colloidal velocity along the active direction, $\langle v_e \rangle$, which is one of the key parameters that determines the Péclet number. For active Brownian particles—used in our simulations—the Péclet number is defined as $Pe= \langle v_e \rangle R_{col}/{D}$, where $D$ denotes the self-diffusion coefficient of the colloid. We have not changed the value of $R_{col}$ and  $\Delta R_{col}$ in this study. As a first step, we have performed simulations for different values of $\theta$, and we adjusted the values of $v_a$ suitably to maintain $\langle v_e \rangle \approx 0.01 ~a_0/\tau_{0}$. This ensured that the Péclet number remained  $ \approx 12$ for different $\theta$, which helped us compare the flow profiles for different $\theta$. These flow profiles are presented in the Results section. We have listed the corresponding values of $v_a$ used in the simulations for each value of $\theta$  in Table~\ref{v_a values}. Note that as we  decrease $\theta$ from $300^{\circ}$ to $60^{\circ}$, we observe that $v_a$ was varied  non-monotonically to keep $ \langle v_e \rangle $ fixed.

\begin{figure}[]
    \centering
    \includegraphics[width=0.85\linewidth]{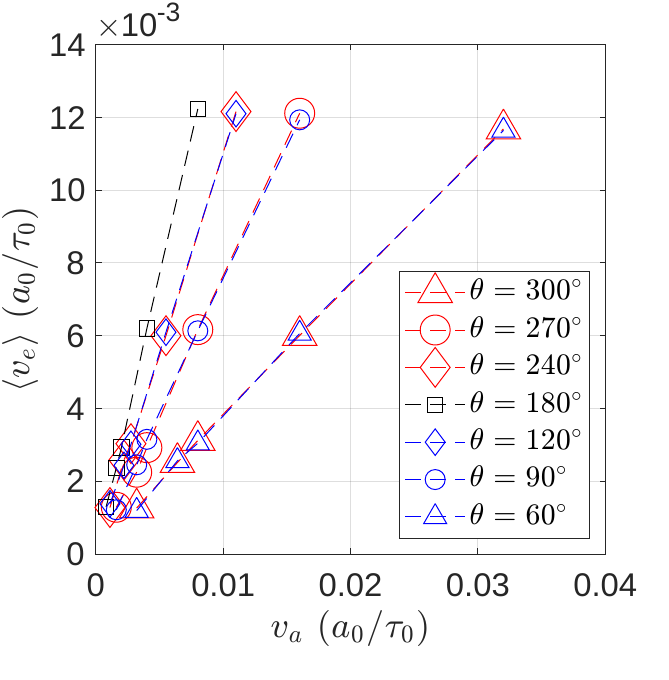}
    \caption{\justifying 
    Mean propulsion velocity $\langle v_e \rangle$ as a function of $v_a$ for different active patch sizes in a $50a \times 50a \times 50a $ simulation box. The patch size is varied through the angle $\theta$. Results are shown for outward pumping at the active surface.} 
    \label{fig:ve_values}
\end{figure}

We also examine how $\langle v_e \rangle$ depends on both $v_a$ and $\theta$. To this end, we perform simulations across a range of $\theta$ and $v_a$ values, computing $\langle v_e \rangle$ in each case. We have presented these results in Fig.~\ref{fig:ve_values}. In Fig.~\ref{fig:ve_values}, we observe that $\langle v_e \rangle$ increases linearly with $v_a$ for a fixed value of $\theta$. Additionally, the value of $\langle v_e \rangle$ is almost equal for both $\theta$ and $2\pi - \theta$ for a given $v_a$, hence exhibiting a non-monotonic variation with respect to $\theta$ (Table~\ref{v_a values}). We have always taken care to ensure that the $\langle v_e \rangle$ is less than $5 \%$ of the sound velocity of the SRD-fluid to avoid artifacts related to the compressibility of the SRD-gas.

We now provide an intuitive argument for the observed non-monotonic dependence of $\langle v_e \rangle$ on $\theta$. The radially outward momentum imparted to the SRD fluid particles can be decomposed into components parallel and perpendicular to $\mathbf{\hat{e}}$. As $\theta$ increases from $45^{\circ}$ to $180^{\circ}$, the number of particles in the upper and lower halves of the active region becomes approximately equal. Consequently, the momentum components perpendicular to $\mathbf{\hat{e}}$ cancel on average. As a result, the net momentum transferred to the fluid saturates with increasing $\theta$ up to $\theta=180^{\circ}$.

For $\theta>180^{\circ}$, partial cancellation also occurs for the momentum component along $\mathbf{\hat{e}}$ because fluid particles are now present both in front of and behind the colloid. Only the remaining uncanceled component along $\mathbf{\hat{e}}$ must be transferred back to the colloid to conserve momentum. With further increase in $\theta$, the number of particles in front of the colloid grows, leading to stronger cancellation between the momentum imparted in the $\mathbf{\hat{e}}$ direction and that imparted in the $-\mathbf{\hat{e}}$ direction to particles behind the colloid. Consequently, the net momentum available to drive the colloid decreases, and larger values of $v_a$ are required to obtain higher $\langle v_e \rangle$.

\begin{figure}[t]
    \includegraphics[width=0.80\linewidth]{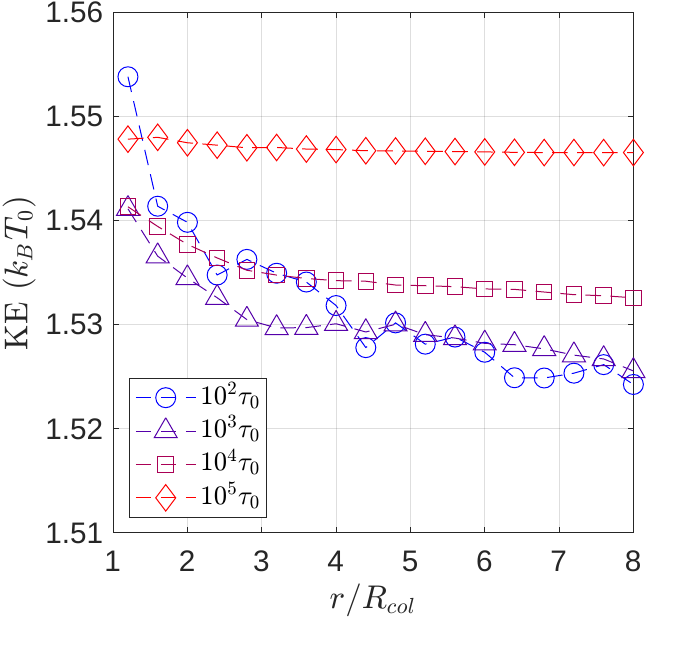}
    \caption{\justifying Kinetic energy (KE) of the fluid in the vicinity of an active colloid with $\theta = 180^\circ$ as a function of normalized distance $r/R_{col}$ from the surface of the colloid. Momentum exchange between the colloid and the MPCD fluid was implemented via hybrid-BC. The fluid KE was computed within concentric spherical shells of thickness $a_0$ centered on the colloid and averaged over different time intervals (see legend).}
    \label{fig:t_shell_actv}
\end{figure}

While the hybrid-BC scheme has previously been shown to maintain the fluid temperature in the absence of a thermostat for a passive colloid, it is important to establish that the same holds for an active colloid. To this end, we perform an analysis analogous to that presented in Fig.~\ref{fig:t_shell_pasv}, but for the active case. The corresponding results are shown in Fig.~\ref{fig:t_shell_actv}. The plot demonstrates that the hybrid-BC scheme is also effective in maintaining a stable fluid temperature for an active colloid.

In summary, in addition to introducing the methods of our simulations, we have also assessed the effectiveness of the newly introduced hybrid-BC and a method to introduce activity. Furthermore, we presented simulation results for a passive spherical colloidal particle subject to the hybrid-BC, establishing that the fluid temperature and the kinetic energy due to the particle's translational and rotational motion are maintained. We have also established the velocity distribution of the active colloid, the effectiveness of the hybrid-BC in maintaining the temperature around the active colloid, and how the colloid's active velocity can be tuned. 

Our simulation box contains $1.25 \times 10^6$ SRD-particles. Moreover, we needed more than $10^7$ SRD iterations and multiple independent runs to calculate the MSD and the time-averaged flow fields. Such calculations are outside the scope of CPU-based HPC clusters. However, the MD+SRD computational scheme is amenable to GPU parallelization using OpenACC in FORTRAN~\cite{tn2023gpu} with minimal code restructuring. Use of Nvidia Tesla V100 GPUs gives a speedup of up to $40$ times compared to running on a single-core CPU, and use of A100 GPUs gives a speedup of up to $60$ times. Statistical quantities for passive colloids are averaged over $100$ independent runs, whereas those for active colloids are averaged over $10$ independent runs. 


\section{Results}
This section is organized into two subsections. In the first subsection, we demonstrate that the coupling of the colloid with the SRD using the Hybrid-BC correctly incorporates the stick boundary condition on the spherical colloid, and thereby, the consequences of incorporating hydrodynamics are observed in  (passive) colloidal dynamics. This is illustrated by the observation of the power law decay of the velocity autocorrelation function of a colloid suspended in the hydrodynamic medium. Furthermore, we establish that the coupling between the colloid and the SRD fluid yields the appropriate hydrodynamic drag on the colloid, as measured by the value of the friction constant $\zeta$ of the sphere. The effectiveness of the no-slip boundary condition can be validated by verifying the Stokes--Einstein relation $ D = k_B T/(6 \pi \eta r) $. We calculate the diffusion constant $D$ from our simulations and show that the viscosity $\eta$ inferred from the Stokes--Einstein relation agrees with the theoretically predicted value of the $\eta$ of SRD-fluid. The theoretical value of $\eta$ is obtained from kinetic theory considerations, and has been previously validated in Poiseuille-flow simulations.

In the second subsection, we investigate flows around an active colloid with inward or outward pumping. The active patch on the particle imparts momentum to a layer of fluid proximal to the colloid and thereby exchanges momentum with the fluid. We also examine how variations in the active patch's surface area modify the induced flow structure and the associated streamlines. We analyze the resulting flow fields around the colloid, some of which exhibit pusher-like or puller-like behavior.


\begin{figure}[]
    \centering

    \begin{subfigure}{0.9\linewidth}
        \centering
        \begin{overpic}[width=0.9\linewidth]{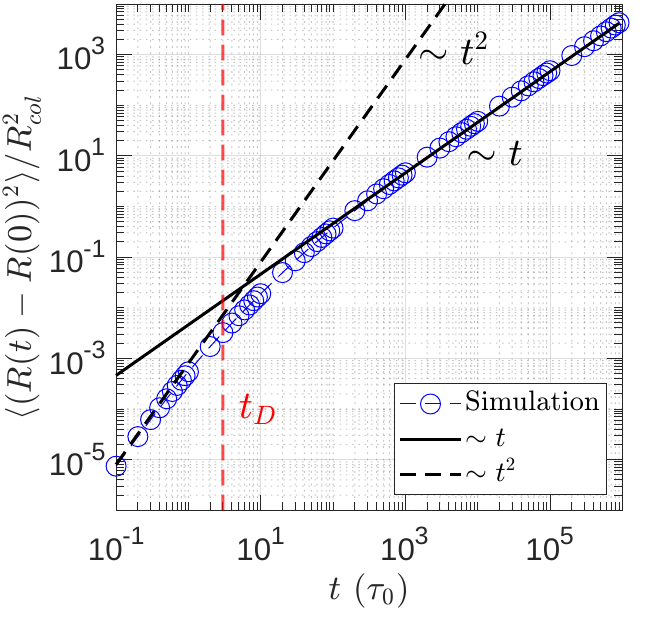}
            \put(40,155){\textbf{(a)}}
        \end{overpic}
        \phantomcaption
        \label{fig:msd_pasv}
    \end{subfigure}
    \hfill
    \begin{subfigure}{0.9\linewidth}
        \centering
        \begin{overpic}[width=0.9\linewidth]{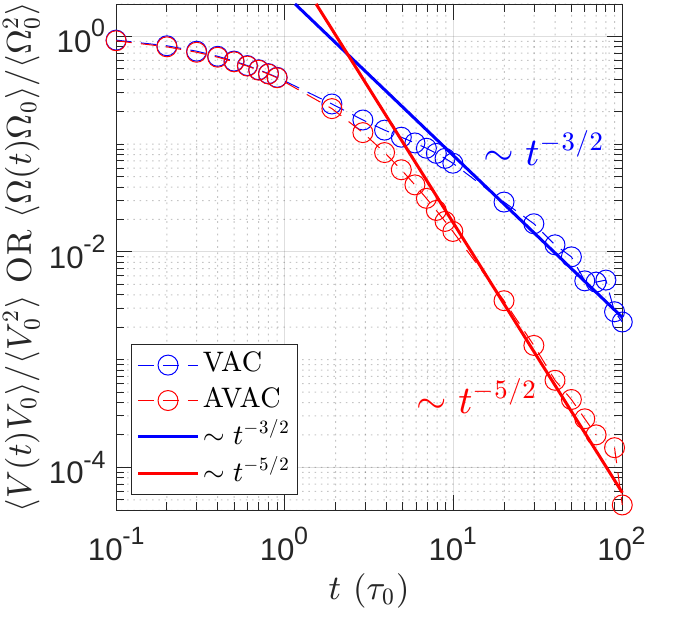}
            \put(140,150){\textbf{(b)}}
        \end{overpic}
        \phantomcaption
        \label{fig:vac_pasv}
    \end{subfigure}
    \caption{\justifying (a) Mean-square displacement (MSD) of a single passive colloid as a function of time ($t/\tau_0$), showing the crossover from short-time ballistic to long-time diffusive behavior. The slope in the diffusive regime yields the friction coefficient $\zeta$ via the Stokes--Einstein relation, and hence the effective fluid viscosity. The solid and dashed black 
    lines are a guide to the eye. 
    The vertical dashed line is at $t_D =M_{col}/\zeta$, and is  the theoretically calculated estimate of the onset of the diffusive regime.
(b) We have plotted the decay of the velocity autocorrelation (VAC) and angular velocity autocorrelation (AVAC) functions with time $t$. At long times, VAC and AVAC exhibit algebraic decay, $\mathrm{VAC} \sim t^{-3/2}$ and $\mathrm{AVAC} \sim t^{-5/2}$. Solid lines denote guides to the eye for the corresponding power laws. The data is averaged over $100$ independent runs.}
    \label{fig:msd_vac}
\end{figure}

\subsection{Passive colloid in SRD fluid with Hybrid BC: validation of Stokes Einstein Relation. }

A colloidal sphere subject to no-slip boundary conditions at its surface experiences a Stokes friction coefficient

\begin{equation}
    \zeta = 6\pi\eta R_{\mathrm{col}}
\end{equation} At thermal equilibrium, the particle obeys the Stokes--Einstein relation, viz.
$D = {k_B T}/{\zeta}$ where   \(D\) is the diffusion coefficient and the friction coefficient is denoted by \(\zeta\). In contrast, a sphere with slip boundary conditions has a reduced Stokes friction coefficient,

\begin{equation}
    \zeta = 4\pi\eta R_{\mathrm{col}}
\end{equation} These well-known results can be used to benchmark and validate the implementation of stick (no-slip) boundary conditions using the hybrid boundary condition. Thereby, we computationally calculate the diffusion constant $D$ of a passive colloid suspended in the SRD hydrodynamic bath and undergoing diffusion. Thereafter, we use the values $R_{col}=2.5~a_0$ and $k_BT=1~k_BT_0$ to calculate the viscosity $\eta$ and find that the deviation from the theoretical value is less than $5\%$.

To elaborate on this aspect further, we calculate the diffusion constant $D$ by the
calculation of the Mean Square Displacement (MSD) of the passive Brownian colloid.  In Fig.~\ref{fig:msd_pasv},  MSD versus time is plotted, where in order to obtain well-averaged MSD  values, we have calculated averages over $100$ independent runs, where each of the simulations was run for $10^6 ~\tau_0$. The MSD shows typical behavior for dilute systems: MSD $\sim t^2$ at short times (ballistic regime) and MSD $\sim t$ at times greater than $t_D$, the time at which diffusive behavior sets in. If we {\em assume} stick boundary conditions, then the onset of the diffusive regime should be at times $t_D = M_{col}/\zeta \approx 3~\tau_0$, which is in agreement with what we see in the data. In the diffusive regime of a colloid, $\langle \Delta r^2 \rangle = 6Dt $. From the slope of the MSD, we find the diffusion coefficient $D$. Using the relation $D = k_BT / 6\pi \eta R_{col}$, we find $\eta = 4.43 ~ m_0/a_0 \tau_0$. This value is very close to the analytical value of $4.62 ~ m_0/a_0 \tau_0$. 


\begin{figure*}[]
    \centering

    \begin{subfigure}{0.24\linewidth}
        \centering
        \begin{overpic}[width=\linewidth]{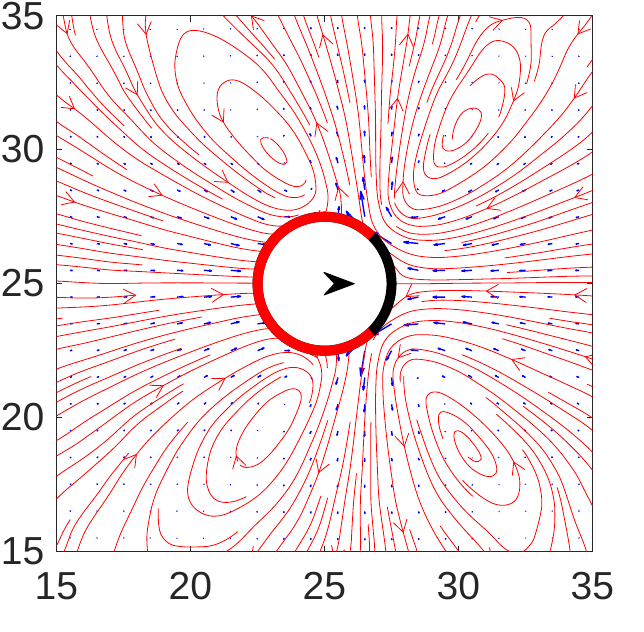}
            \put(15,105){\colorbox{white}{\textbf{(a)}}}
            \put(86,20){\colorbox{white}{\textbf{$270^{\circ}$}}}
        \end{overpic}
    \end{subfigure}
    \hfill
    \begin{subfigure}{0.24\linewidth}
        \centering
        \begin{overpic}[width=\linewidth]{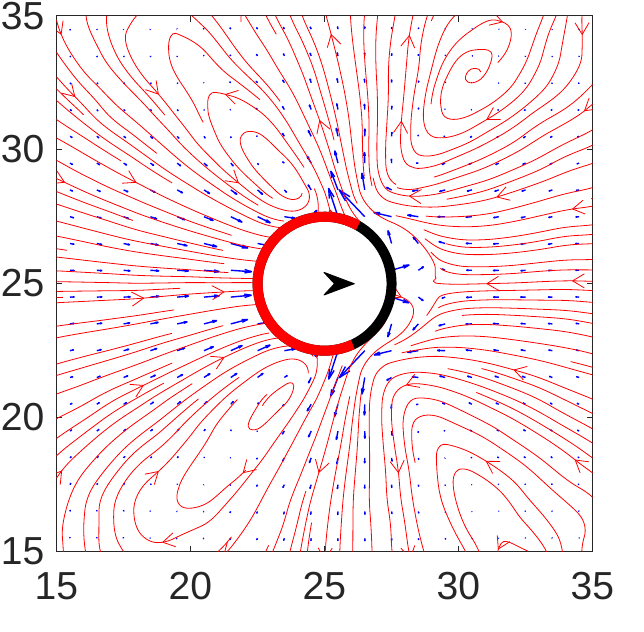}
            \put(15,105){\colorbox{white}{\textbf{(b)}}}
            \put(86,20){\colorbox{white}{\textbf{$240^{\circ}$}}}
        \end{overpic}
    \end{subfigure}
    \hfill
    \begin{subfigure}{0.24\linewidth}
        \centering
        \begin{overpic}[width=\linewidth]{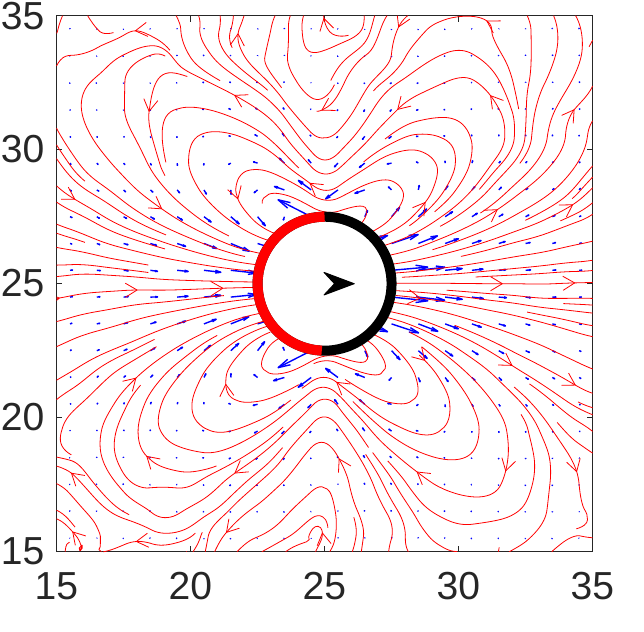}
            \put(15,105){\colorbox{white}{\textbf{(c)}}}
            \put(86,20){\colorbox{white}{\textbf{$180^{\circ}$}}}
        \end{overpic}
    \end{subfigure}
    \hfill
    \begin{subfigure}{0.24\linewidth}
        \centering
        \begin{overpic}[width=\linewidth]{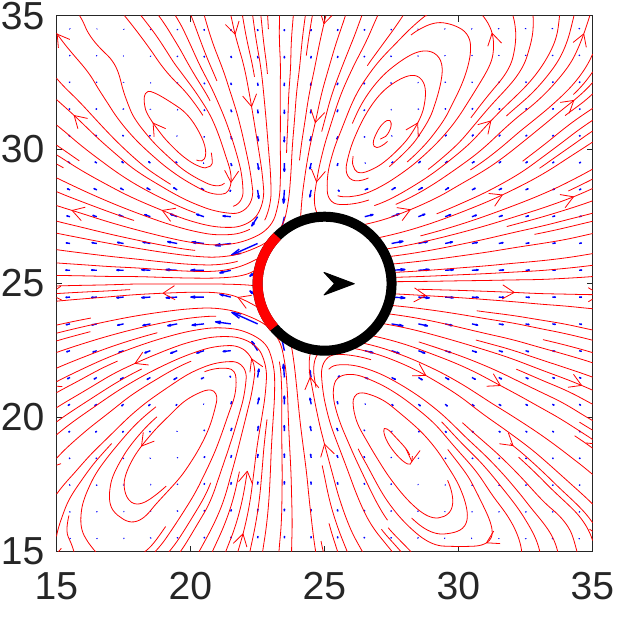}
            \put(15,105){\colorbox{white}{\textbf{(d)}}}
            \put(91,20){\colorbox{white}{\textbf{$90^{\circ}$}}}
        \end{overpic}
    \end{subfigure}
    \hfill
    \begin{subfigure}{0.24\linewidth}
        \centering
        \begin{overpic}[width=\linewidth]{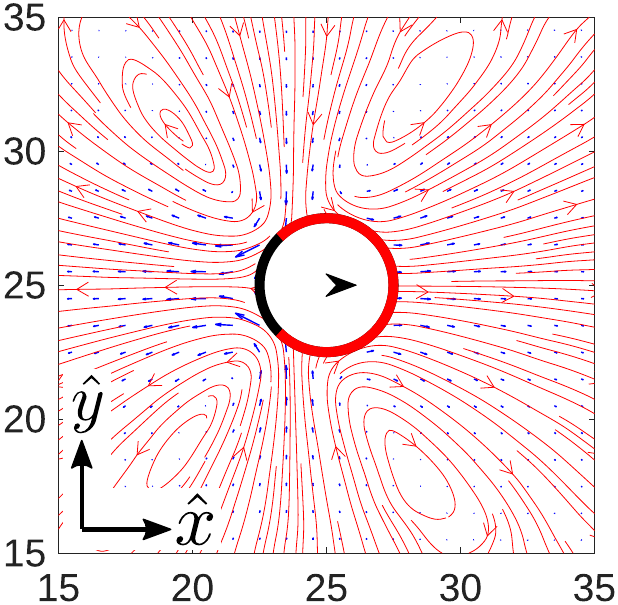}
            \put(15,105){\colorbox{white}{\textbf{(e)}}}
            \put(93,15){\colorbox{white}{\textbf{$270^{\circ}$}}}
        \end{overpic}
    \end{subfigure}
    \hfill
    \begin{subfigure}{0.24\linewidth}
        \centering
        \begin{overpic}[width=\linewidth]{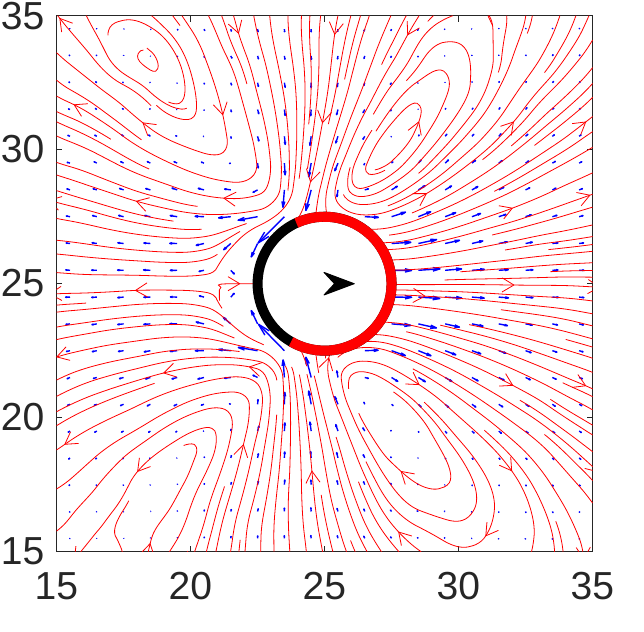}
            \put(15,105){\colorbox{white}{\textbf{(f)}}}
            \put(86,20){\colorbox{white}{\textbf{$240^{\circ}$}}}
        \end{overpic}
    \end{subfigure}
    \hfill
    \begin{subfigure}{0.24\linewidth}
        \centering
        \begin{overpic}[width=\linewidth]{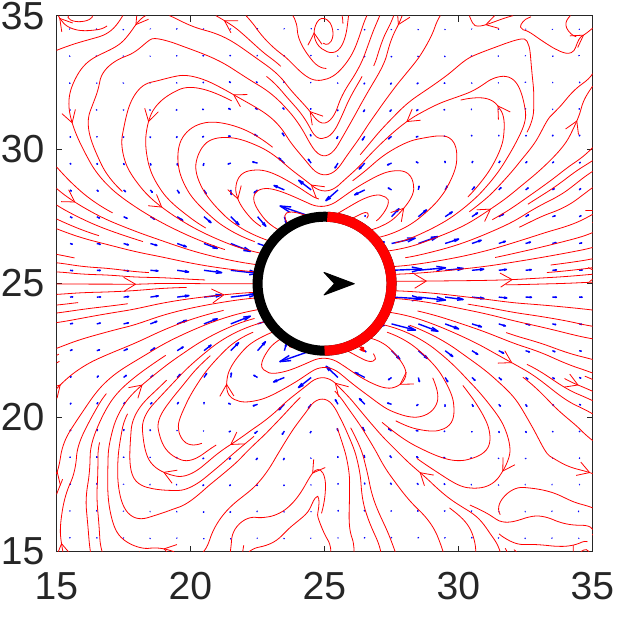}
            \put(15,105){\colorbox{white}{\textbf{(g)}}}
            \put(86,20){\colorbox{white}{\textbf{$180^{\circ}$}}}
        \end{overpic}
    \end{subfigure}
    \hfill
    \begin{subfigure}{0.24\linewidth}
        \centering
        \begin{overpic}[width=\linewidth]{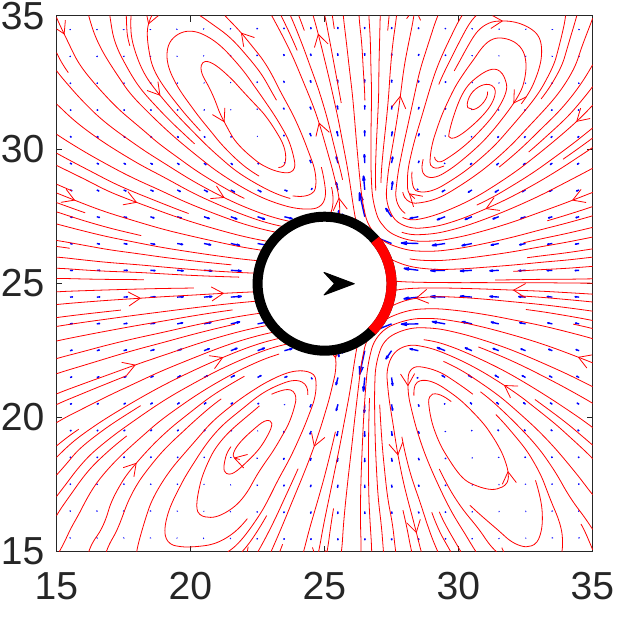}
            \put(15,105){\colorbox{white}{\textbf{(h)}}}
            \put(91,20){\colorbox{white}{\textbf{$90^{\circ}$}}}
        \end{overpic}
    \end{subfigure}
    \caption{\justifying Flow profiles around an active colloid for different values of $\theta$.  Here, blue arrows show the fluid flow fields and red lines show streamlines. The top row corresponds to outward pumping and the bottom row to inward pumping. The black arrow denotes the active direction $\mathbf{\hat{e}}$, which is aligned with $\mathbf{\hat{x}}$ for analysis. The arc in red shows the active region, and the black arc shows the inert region. The Columns from left to right show $\theta = 270^\circ$, $240^\circ$, $180^\circ$, and $90^\circ$. For $\theta > 180^\circ$, the flow is puller-like for outward pumping and pusher-like for inward pumping; for $\theta < 180^\circ$, this behavior is reversed. At $\theta = 180^\circ$, the flow field is qualitatively akin to that of a neutral swimmer of the squirmer model, for both pumping schemes. The direction indicated by the arrow defines the front of the colloid. The colloid center is located at the center ($25a, 25a, 25a$) of a $50a \times 50a \times 50a$ simulation box with periodic boundary conditions.  Flow profiles are averaged over $10$  independent runs. The fluid velocities are calculated in the shifted and reoriented frame, refer to {\em Calculation of streamlines} in the text. In this frame of reference, we average over the particles sandwiched between the $ z/a=24$ and $z/a=25$ planes to obtain the fluid velocity fields. Flow profiles for $\theta= 300^{\circ}$,$120^{\circ}$ and $60^{\circ}$ are given in the supplementary information (SI)~\cite{Supplemental}.}
    \label{fig:actv_flow_profile}
\end{figure*}
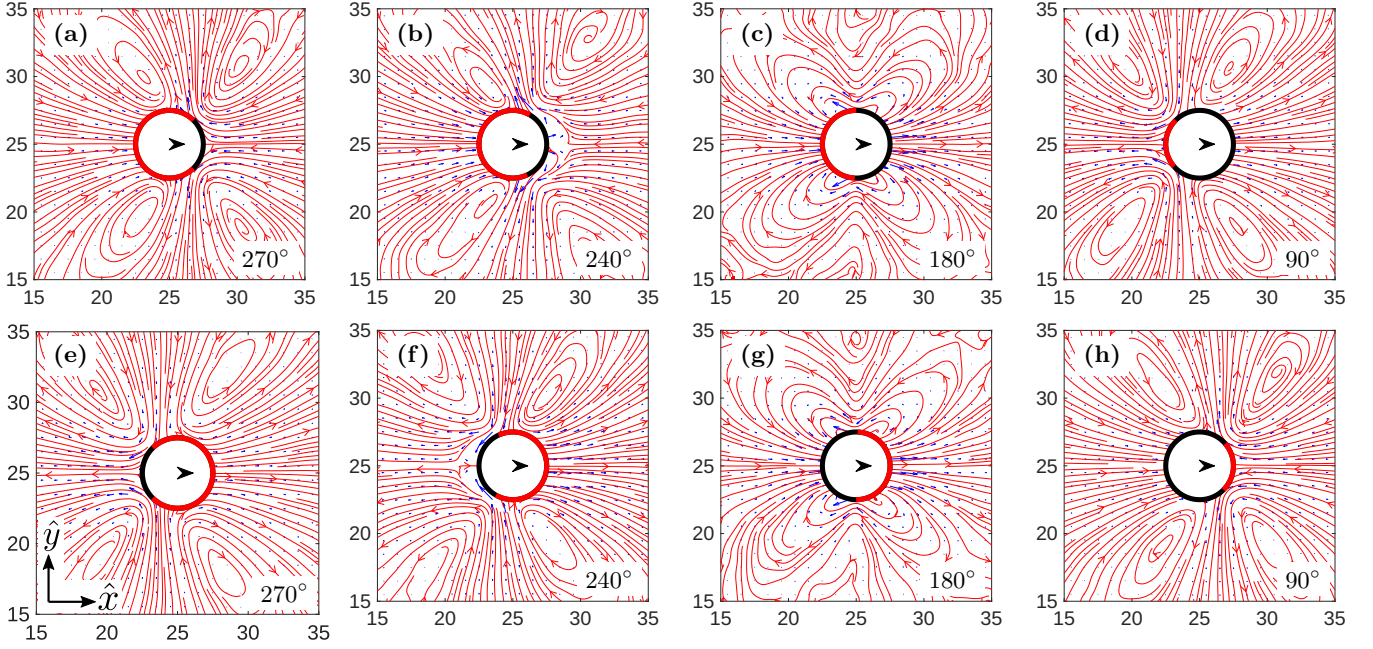

Thus, we conclude that our Hybrid boundary condition mimics the no-slip boundary condition rather accurately. Previously, we also showed that hybrid-BC maintains the correct temperature of the fluid, which is proximal to the colloid surface. Moreover, hybrid-BC maintains the appropriate distribution of the translational and angular speeds of the colloid at temperature $k_BT_0$, without the use of an additional thermostat, as seen in Fig.~\ref{fig:mba}.

We also calculate the translational and angular velocity autocorrelation functions (VAC and AVAC) to assess hydrodynamic effects on the passive colloid. In the absence of hydrodynamic interactions, the Langevin equation predicts exponential decay of both correlations. In contrast, hydrodynamic coupling leads to long-time algebraic tails, with VAC and AVAC decaying as $t^{-3/2}$ and $t^{-5/2}$, respectively~\cite{padding2005stick}. Fig.~\ref{fig:msd_vac} shows this behavior over approximately one decade for times greater than $10 \tau_0$. Note that the kinematic viscosity is $\eta/\rho = 0.46 ~ a_0^2/ \tau_0$ ($\rho = m \gamma$) which  sets the time scale for viscous momentum diffusion through the fluid. The long-time tails in the VAC and AVAC emerge after approximately $20$ times this scale.

\subsection{Flow fields around active colloid}

We define the central axis as the line passing through the colloid center and aligned with the orientation vector. The lateral axis denotes any line passing through the colloid center and perpendicular to the central axis. For ease of communication, we refer to the plane passing through the center of the colloid, with the orientation vector normal to it, as the equatorial plane. Correspondingly, we refer to the points where the central axis intersects the colloidal surface as the poles.  

{\em Calculation of streamlines:} We analyze the flow profiles generated by active colloids as the area of the catalytic patch is varied through changes in $\theta$. To compute the fluid velocity field around the colloid, we divide the simulation box into cubic cells of edge length $a$. For each cell, we calculate the instantaneous local velocity by averaging the velocities of all SRD particles within that cell. Consequently, time averaging yields the fluid velocity field at a spatial resolution set by the cell size. The resulting velocity field provides the space-dependent flow profiles around the colloid.

The position and orientation of the colloidal particle change with time due to diffusion. Consequently, special care must be taken when time-averaging the fluid flow profiles. For the calculation of the flow profiles shown in Fig.~\ref{fig:actv_flow_profile}, we shift and rotate the coordinate frame at every iteration prior to computing the flow profile. Specifically, we first translate the frame so that the colloid center coincides with the center of the simulation box, and then rotate it such that the $x$ axis aligns with the orientation vector $\mathbf{\hat{e}}$. We subsequently transform the positions and velocities of the SRD particles into this reference frame before performing the averaging.

Fig.~\ref{fig:actv_flow_profile} shows fluid flow profiles around active colloid 
for different $\theta$. We show data for $\theta = 270^{\circ},240^{\circ},180^{\circ}$ and $90^{\circ}$ as we move from left to right.   The top row shows data with outward pumping, whereas the bottom row shows data with inward pumping. For different values of $\theta$, we adjust the value of $v_a$ to maintain $\langle v_e \rangle \approx 0.01~a_0/\tau_0$. To calculate these flow profiles, we have averaged over $10$ independent runs, each of length $10^6~ \tau_0$. Before collecting data to calculate flow profiles, we allowed the system to reach steady state over $10^4 ~\tau_0$.

{\em Crossover from Pusher to Puller:} For outward pumping, we observe that for $\theta = 270^{\circ}$ and $240^{\circ}$ the flow profiles are puller-like, whereas for $\theta = 90^{\circ}$ the flow profile is pusher-like, as clarified in the next paragraph. For $\theta = 180^{\circ}$, the flow field resembles that of a neutral swimmer of the squirmer model. In contrast, for inward pumping, we obtain puller-like flow profiles for $\theta = 90^{\circ}$ and pusher-like flow profiles for $\theta = 270^{\circ}$ and $240^{\circ}$. Similar to the outward pumping case, the flow field for $\theta = 180^{\circ}$ qualitatively corresponds to that of a neutral swimmer of the squirmer model. We have also calculated flow profiles for $\theta = 300^{\circ},120^{\circ}$ and $60^{\circ}$ for both outward and inward pumping. This is shown in the Supplementary Information, Section-1. 

We clarify that a pusher-like flow is characterized by outward fluid flow along the central axis and inward flow toward the colloidal surface in the lateral directions near the equatorial plane. Conversely, in a puller-like flow, fluid is drawn inward along the central axis and flows outward along the lateral directions near the equatorial plane. The puller-like flow profile, even with outward pumping, and the pusher-like flow profile with inward pumping are non-intuitive a priori. Thereby, we explain the underlying mechanism for the transition from pusher to puller as we change $\theta$,  even as $v_e=0.01 ~ a_0 /\tau_0$ is kept unchanged in our simulations by adjusting $v_a$.

We expect that the transition from a puller to a pusher for outwards pumping (or vice-versa for inwards pumping) should occur near $\theta = 180^{\circ}$, although we will systematically investigate this in the future. In summary, the proposed scheme of introducing  activity on the surface of the colloid can be used to model both types of swimmers (Pusher and Puller), by varying $\theta$, and is independent of whether colloids have outward  or inward pumping
at the active surface.

In our model, activity is implemented by imparting additional momentum to fluid particles in the vicinity of the colloidal surface, which can be interpreted as an effective active force acting on the surrounding fluid. For small values of $\theta$ (e.g., $\theta = 90^\circ$), this forcing is predominantly localized near the central axis.

\begin{figure}[]
    \centering

    \begin{subfigure}{0.49\linewidth}
        \centering
        \begin{overpic}[width=\linewidth]{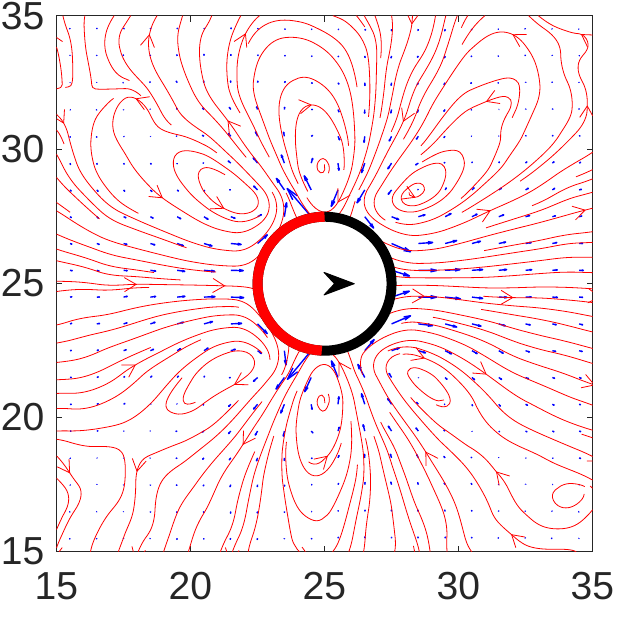}
            \put(12,105){\colorbox{white}{\textbf{(a)}}}
        \end{overpic}
        \phantomcaption
        \label{fig:static_xy}
    \end{subfigure}
    \hfill
    \begin{subfigure}{0.49\linewidth}
        \centering
        \begin{overpic}[width=\linewidth]{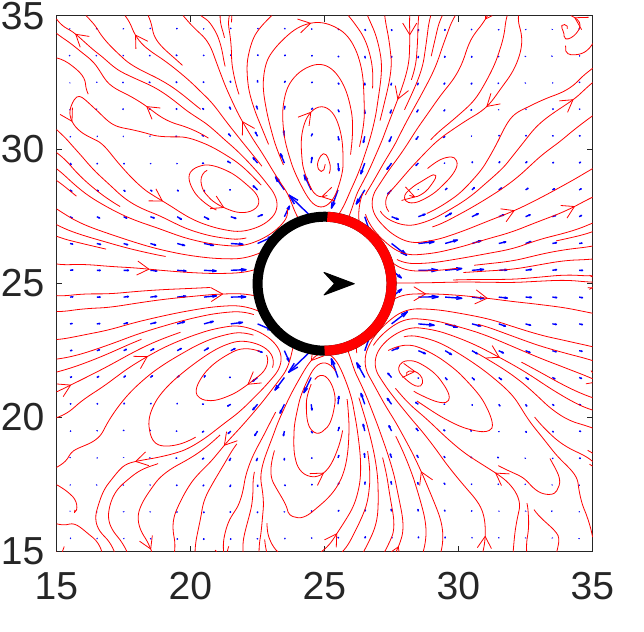}
            \put(12,105){\colorbox{white}{\textbf{(b)}}}
        \end{overpic}
        \phantomcaption
        \label{fig:static_yz}
    \end{subfigure}

    \caption{\justifying Flow profiles for an active colloid with $\theta = 180^\circ$ for (a) inward and (b) outward pumping, with the colloid held fixed in space. Here, blue arrows show the flow field and red lines show streamlines.}
    \label{fig:static_flow_profile}
\end{figure}

{\em Active patch with $\theta = 90^{\circ}$}: For the outward-pumping colloid, the active force is applied primarily at the rear of the colloid, driving fluid away from the surface in this region. Simultaneously, the colloid's forward motion transfers momentum to the surrounding fluid, reinforcing the outward flow at its front. Owing to the incompressibility of the SRD fluid, this axial outflow at the poles is compensated by an inward flux along the lateral directions, resulting in flow toward the colloid in regions close to the equatorial plane.

In contrast, for the inward-pumping case, the active force acts predominantly at the front of the colloid, drawing fluid toward the surface. At the same time, the forward motion of the colloid entrains fluid from the rear, producing inward flow there as well. Because we ensure that the values of $v_a$ are low enough to ensure that the fluid flow remains incompressible, this axial inflow is balanced by an outward flux along the lateral directions. This leads to fluid flow away from the colloidal surface at the equatorial plane.

{\em Larger values of $\theta$:} As we increase $\theta$ to increase the size of the active patch, the fluid region over which the active force acts extends progressively toward the lateral axes. In the outward-pumping case, forcing along the lateral directions generates fluid motion away from the colloid along the lateral axis. Interestingly, the outward flow in the lateral direction results in a net inward flow along the central axis due to the fluid's incompressibility. This occurs despite the presence of radially outward forces acting on the fluid. Conversely, in the inward-pumping case, forcing in the lateral directions induces fluid motion toward the colloidal surface, leading to an outward flow along the central axis. For outward pumping active colloids with $\theta > 180^{\circ}$, the component of active forces along the central axis directions are opposite on the two sides of the equatorial plane. However, the force along the lateral directions does not cancel out near the colloidal surface at the equatorial plane. A similar scenario is also true for inward pumping active colloids with $\theta > 180^{\circ}$

\begin{figure}[t]
    \centering

    \begin{subfigure}{0.49\linewidth}
        \centering
        \begin{overpic}[width=\linewidth]{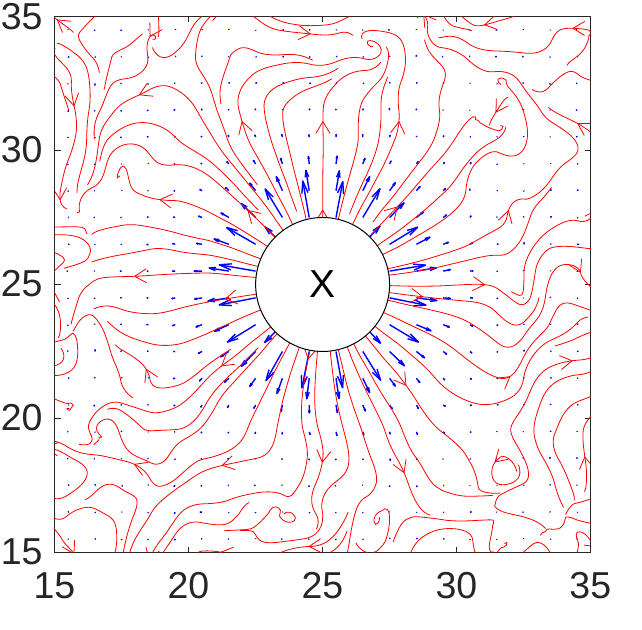}
            \put(15,105){\colorbox{white}{\textbf{(a)}}}
        \end{overpic}
        \phantomcaption
        \label{fig:pusher_90_24}
    \end{subfigure}
    \hfill
    \begin{subfigure}{0.49\linewidth}
        \centering
        \begin{overpic}[width=\linewidth]{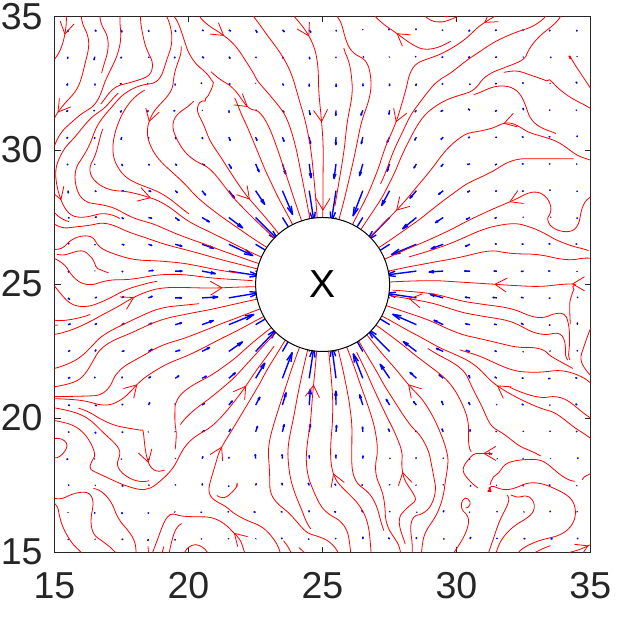}
            \put(15,105){\colorbox{white}{\textbf{(b)}}}
        \end{overpic}
        \phantomcaption
        \label{fig:pusher_90_26}
    \end{subfigure}
    \begin{subfigure}{0.49\linewidth}
        \centering
        \begin{overpic}[width=\linewidth]{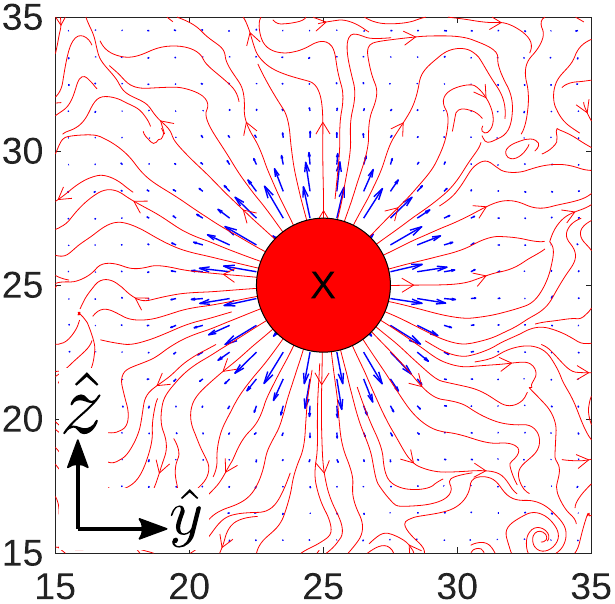}
            \put(15,105){\colorbox{white}{\textbf{(c)}}}
        \end{overpic}
        \phantomcaption
        \label{fig:puller_90_24}
    \end{subfigure}
    \hfill
    \begin{subfigure}{0.49\linewidth}
        \centering
        \begin{overpic}[width=\linewidth]{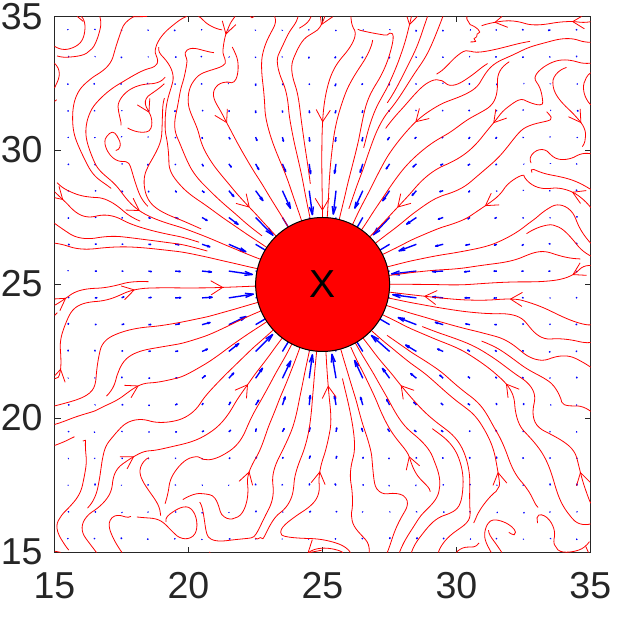}
            \put(15,105){\colorbox{white}{\textbf{(d)}}}
        \end{overpic}
        \phantomcaption
        \label{fig:puller_90_26}
    \end{subfigure}

    \caption{\justifying Fluid flow profiles around an active colloid with $\theta = 180^\circ$ in the $y$--$z$ plane, viewed along the $-\mathbf{\hat{e}}$ direction. Blue arrows show flow field and red lines show streamlines. The active direction $\mathbf{\hat{e}}$ is normal to the plane and points out of the page. The top row corresponds to outward pumping and the bottom row to inward pumping. Panels (a) and (c) correspond to the plane $x/a = 24.5$, while panels (b) and (d) correspond to $x/a = 25.5$.}
    \label{fig:flow_yz}
\end{figure}


{\em Static active colloid:} We also emphasize here that instead of shifting and rotating the whole frame to calculate the flow profile, one can also calculate the flow profile by keeping the active colloid fixed at one point. This can be done by not updating the position and orientation of the active colloid. While this approach significantly simplifies the computation of flow profiles, it does not yield physically accurate results. In Fig.~\ref{fig:static_flow_profile}, we present the flow field obtained using this procedure for $\theta = 180^\circ$ in both inward and outward pumping cases. The resulting profiles differ distinctly from those obtained by shifting and rotating the whole frame, as shown in Fig.~\ref{fig:actv_flow_profile}.


\begin{figure}[t]
    \centering

    \begin{subfigure}{0.95\linewidth}
        \centering
        \begin{overpic}[width=0.85\linewidth]{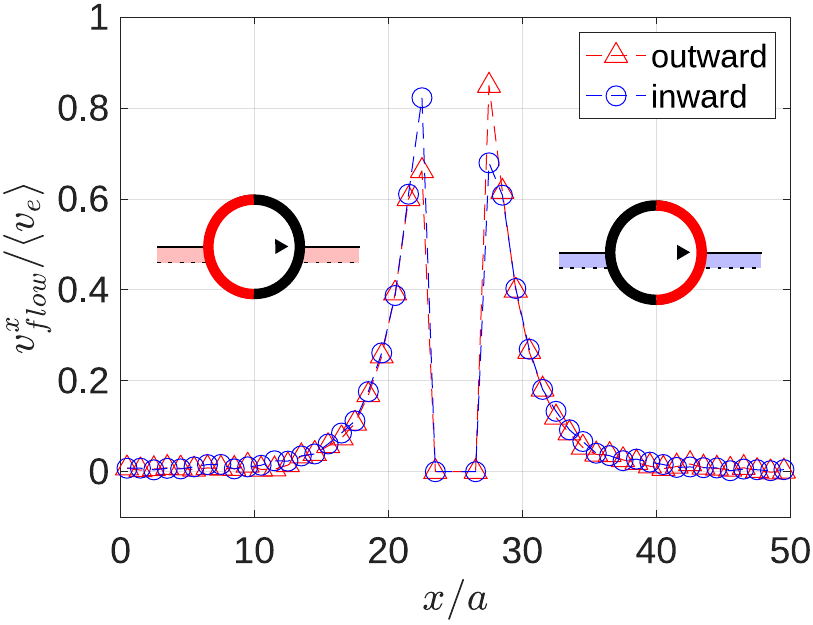}
            \put(40,135){\textbf{(a)}}
        \end{overpic}
        \phantomcaption
        \label{fig:x_dir_magnitude}
    \end{subfigure}
    \hfill
    \begin{subfigure}{0.95\linewidth}
        \centering
        \begin{overpic}[width=0.85\linewidth]{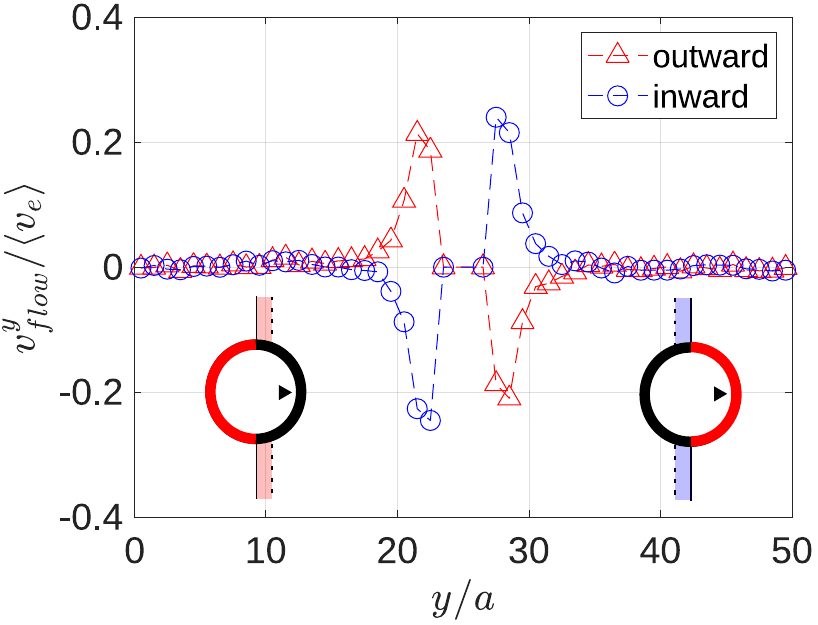}
            \put(40,135){\textbf{(b)}}
        \end{overpic}
        \phantomcaption
        \label{fig:y_dir_magnitude}
    \end{subfigure}
    \caption{\justifying (a) We show the x-component of the velocity $v^x_{flow}$ along the line passing next to the center of the colloid in $\mathbf{\hat{e}}$ direction. We have shown the variation in $v^x_{flow}$ along the $x$. Details of the procedure for calculating fluid velocity are provided in the text; a schematic is included in the figure to help the reader understand the details. The fluid velocity is normalized by the magnitude of the mean velocity of the active colloid $\langle v_e \rangle$. We show data for both {\em outwards} and {\em inwards} pumping. (b) We also show the $y$ component of the velocity of fluid $v^{y}_{flow}$ (suitably normalized by $\langle v_e \rangle$) along a line in the $\mathbf{\hat{y}}$ direction, i.e., perpendicular to  $\mathbf{\hat{e}}$ along a line next to the lateral axis. Note that the fluid velocity inside the sphere is marked as zero as there is no fluid inside.}
    \label{vel_prof_theta_is_180}
\end{figure}

{\em Streamlines in the lateral directions:} We now examine the flow fields in the plane perpendicular to $\mathbf{\hat{e}}$. Since $\mathbf{\hat{e}}$ is aligned along the $x$-direction, this corresponds to the $yz$ plane. In Fig.~\ref{fig:flow_yz}, we present the flow fields in the $yz$ plane for $\theta = 180^\circ$. The top row shows the outward-pumping case, while the bottom row corresponds to inward pumping. In all panels, the vector $\mathbf{\hat{e}}$ is directed out of the plane of the page.

For Fig.~\ref{fig:flow_yz}(a) and (c), the flow field is obtained by averaging the SRD particle velocities over the slab $24 \leq x/a \leq 25$, corresponding to the plane at $x/a = 24.5$. Similarly, for Fig.~\ref{fig:flow_yz}(b) and (d), the averaging is performed over $25 \leq x/a \leq 26$, corresponding to the plane at $x/a = 25.5$. Here, the colloid center is located at $(25a, 25a, 25a)$. In these streamlines, one can clearly observe the expected radial symmetry in the $yz$ plane.

{\em Quantitative measurements of fluid flows:} From the streamlines of flow presented above, it is not possible to decipher the magnitude of the velocities of fluid flow close to the active colloidal surface, especially relative to the active colloid velocity. To that end, we have plotted the component of the fluid velocities, suitably normalized by the mean speed of the colloid $\langle v_e \rangle$, along $\mathbf{\hat{x}}$, that is along the central axis.  Fig.~\ref{vel_prof_theta_is_180}(a) shows data for the swimmer with $\theta =180^{\circ}$ for both outward pumping and inwards pumping.  In Fig.~\ref{vel_prof_theta_is_180}(b), we have also plotted the normalized fluid velocity component along $\mathbf{\hat{y}}$ along a line passing through the colloid center, i.e., along one of the lateral axes. 

\begin{figure}[t]
    \centering

    \begin{subfigure}{0.8\linewidth}
        \centering
        \begin{overpic}[width=\linewidth]{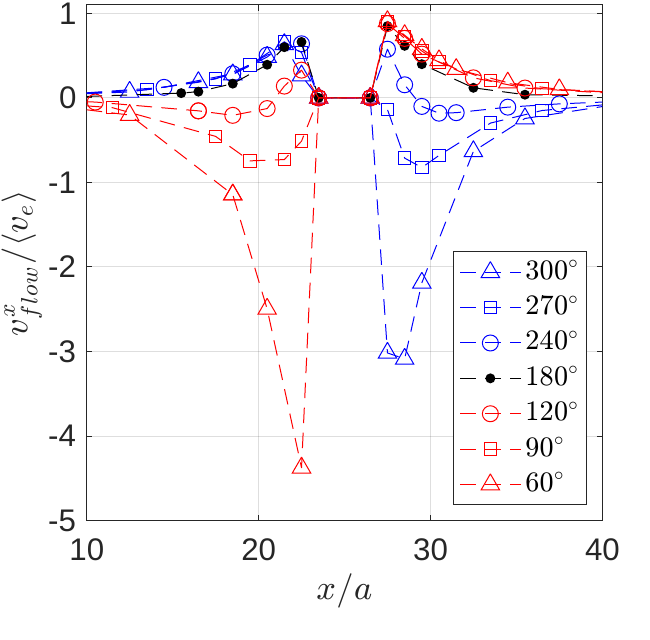}
            \put(35,172){\textbf{(a)}}
        \end{overpic}
        \phantomcaption
        \label{fig:push_x}
    \end{subfigure}
    \hfill
    \begin{subfigure}{0.98\linewidth}
        \centering
        \begin{overpic}[width=0.8\linewidth]{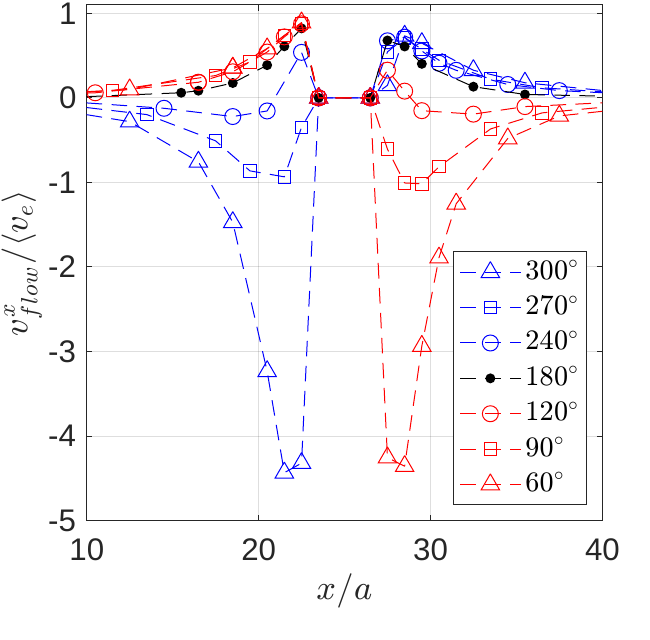}
            \put(35,172){\textbf{(b)}}
        \end{overpic}
        \phantomcaption
        \label{fig:pull_x}
    \end{subfigure}

    \caption{\justifying We show the $x$-component of the velocity $v^x_{flow}$ along the line passing next to the center of the colloid in $\mathbf{\hat{e}}$ direction for different values of $\theta$. The values of $\theta$ are given in the legend. We have shown the variation in $v^x_{flow}$ along the $x$.  Note that the fluid velocity inside the sphere is marked as zero, as there is no fluid inside. The top figure labeled (a) shows the flow magnitude for the {\em outward} pumping case, and the bottom figure (b) shows the flow magnitude for the {\em inward} pumping case.  The procedures for fluid velocity calculation are the same as in Fig.~\ref{vel_prof_theta_is_180}(a).}
    \label{fig:magnitude_x}
\end{figure}

To calculate the mean fluid velocity component along $\mathbf{\hat{e}}$, we take the fluid particles sandwiched between the planes $ y/a=24$, $y/a=25$ and the planes $z/a= 24$,  $z/a=25$ and calculate time-averaged mean values of $v^x_{flow}(x)$ at different values of $x$. Similarly, to calculate the velocity of fluid $v^y_{flow} (y)$, we time average the velocity of the SRD-particles sandwiched between the planes $x/a=25$, $x/a=26$, and the planes $z/a=25$, $z/a=26$ for outward pumping. The velocities are averaged over SRD particles, which are proximal to the passive part of the colloid. For the inward pumping case, the colloid moves so that the active surface is in the forward direction. Thereby, we calculate the local fluid velocities of particles which are sandwiched between the planes  $x/a=24$, $x/a=25$, and the planes $z/a=24$, $z/a=25$. The regions which are being used to calculate fluid velocities are shown by schematics in Fig.~\ref{vel_prof_theta_is_180}.

\begin{figure}[t]
    \centering

    \begin{subfigure}{1.0\linewidth}
        \centering
        \begin{overpic}[width=0.8\linewidth]{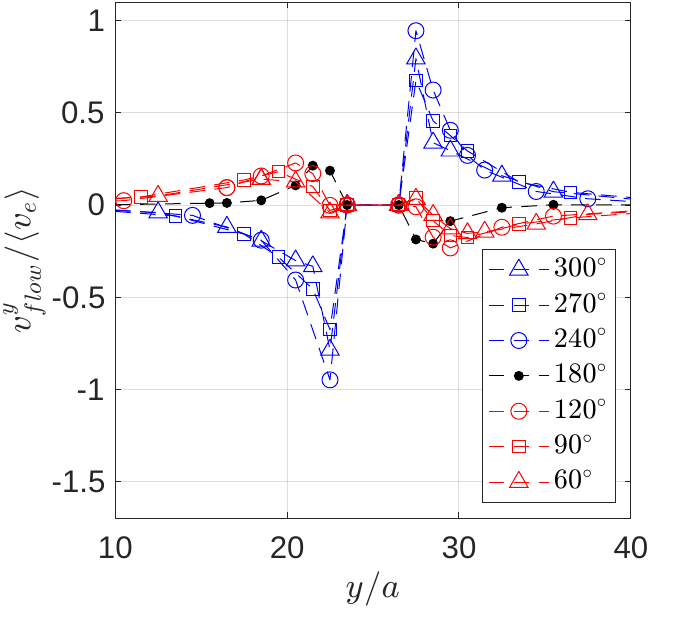}
            \put(45,170){\textbf{(a)}}
        \end{overpic}
        \phantomcaption
        \label{fig:push_y}
    \end{subfigure}
    \hfill
    \begin{subfigure}{1.0\linewidth}
        \centering
        \begin{overpic}[width=0.8\linewidth]{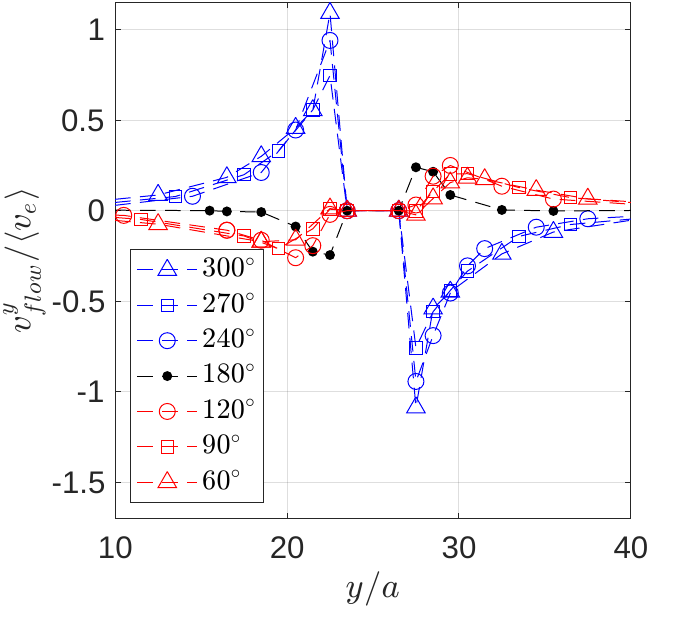}
            \put(45,170){\textbf{(b)}}
        \end{overpic}
        \phantomcaption
        \label{fig:pull_y}
    \end{subfigure}

    \caption{\justifying We also show the $y$ component of the velocity of fluid $v^{y}_{flow}$ along a line in the $\mathbf{\hat{y}}$ direction, i.e. perpendicular to  $\mathbf{\hat{e}}$ along a line next to the lateral axis. Data is plotted for different values of $\theta$. The values of $\theta$ are given in the legend. Note that the fluid velocity inside the sphere is marked as zero, as there is no fluid inside. The top figure labeled (a) shows the flow magnitude for the {\em outward} pumping case, and the bottom figure (b) shows the flow magnitude for the {\em inward} pumping case. The procedures for fluid velocity calculation are the same as in Fig.~\ref{vel_prof_theta_is_180}(b).}
    \label{fig:magnitude_y}
\end{figure}

From the symmetry of the problem, we expect the flow in the $\mathbf{\hat{z}}$ direction along the line passing through the center of the sphere to be the same as in the $\mathbf{\hat{y}}$ direction. In Fig.~\ref{vel_prof_theta_is_180}(a), $v^x_{\mathrm{flow}}$ just in front of the colloid is larger for outward pumping than for inward pumping. For inward pumping, the active patch lies at the front as the colloid moves along $\mathbf{\hat{e}}$, and momentum is imparted to the fluid opposite to the direction of motion. Nevertheless, the fluid velocities immediately in front of and behind the colloid remain comparable to the colloid speed. As expected, the flow weakens with increasing distance from the colloidal surface. The fluid flows in the lateral directions, just ahead and trailing the mid-plane (refer to Fig.~\ref{vel_prof_theta_is_180}(b)), show $v^y_{flow}/v_e < 0.5$, and the relative values for this for inward and outward pumping are decided by the net flows in the $\mathbf{\hat{e}}$ directions.

We show the fluid velocities along the central axis $\mathbf{\hat{x}}$ and a lateral axis $\mathbf{\hat{y}}$ in Fig.~\ref{fig:magnitude_x} and Fig.~\ref{fig:magnitude_y}, respectively, for active colloids with different values of $\theta$. To calculate the fluid velocities and the choice of the axes' positions, we follow the procedure described for Fig.~\ref{vel_prof_theta_is_180} and shown in the schematics. The plots at the top of Fig.~\ref{fig:magnitude_x} and Fig.~\ref{fig:magnitude_y} are for outward pumping, and the bottom plots are for inward pumping.
   
In Fig.~\ref{fig:magnitude_x}(a), we can resolve the strength of flow along the central axis at the front and rear of the colloid for outward pumping. For $ \theta= 60^\circ$,  momentum is imparted to SRD particles near a small patch at the rear of the colloid, causing outward radial motion of the fluid but maintaining $v_e = 0.012~ a_0/\tau_0$. Thereby, there is a net fluid flow in the $-\mathbf{\hat{x}}$, and correspondingly, we observe a relatively large value of $v^x_{flow}/v_e \approx 4$ at the back of the colloid. Fluid flows in from the lateral directions, and the fluid is again pushed out towards $\mathbf{\hat{x}}$ in front of the colloid. Consequently, the  $\theta= 60^\circ$ colloid is a pusher.

On the other hand, momentum is imparted to the fluid radially outwards over a large fraction of the surface of the colloid for a colloid with $\theta = 300^{\circ}$. As a result, the fluid is transported away from the active surface, and the active surface covers a large fraction of the colloid's surface. Since the simulation parameters are chosen so that the fluid can be considered incompressible, the fluid must flow in from the front, where the passive part of the surface is situated, and hence we see a large negative value of $v^x_{flow}/v_e \approx -3$. Fluid flows out from the lateral directions, and moreover, fluid at the rear of the moving colloid flows towards the surface, but with $v^x_{flow}/v_e <1$. Therefore, the $\ theta=300^\circ$ colloid is a puller. Throughout, the fluid velocities remain below $0.05$ times the speed of sound of the SRD fluid as $v_e$ is always maintained at $v_e = 0.012,a_0/\tau_0$. The ensures incompressible flow. For all different values of $\theta$ considered except $\theta = 180^{\circ}$, the direction of flow is reversed from the rear to the front, though the flow speeds can have different values.

Figures Fig.~\ref{fig:magnitude_x}(b) and Fig.~\ref{fig:magnitude_y}(b) show $v_{flow}^x / v_e$ versus $x$ and $v_{flow}^y / v_e$ versus $y$ along lines passing through the center of the colloid along the lateral (vertical) direction. For fluid velocity calculations and choice of axes, we followed the same conventions that have been described in detail for Fig.~\ref{vel_prof_theta_is_180} (b) for outward and inward pumping, respectively. For all the cases that we observed, the direction of flow is opposite at the top and bottom of the colloid. The analysis is analogous to the previous two paragraphs, except that the magnitudes and directions of flow along $ \mathbf{\hat{y}}$ (and $ \mathbf{\hat{z}}$) are determined by mass conservation and by flows along the active directions.

\section{Conclusion}

We have presented a novel computer-simulation framework for generating a wide range of hydrodynamic flow fields around a spherical colloid. The proposed scheme is motivated by experimental studies of synthetic patchy active colloids~\cite{rohde2025programmable}, in which chemical reactions near the active patch drive the surrounding fluid. Depending on the nature of the activity, the fluid may be driven either toward or away from the colloidal surface. Our simulation model is designed to capture these two mechanisms through two distinct modes of activity: outward pumping and inward pumping. We systematically investigate both modes for different active patch sizes. Our results demonstrate that patch size significantly influences the resulting flow profiles in both cases. By appropriately tuning the active patch area, the active colloid in our simulations can generate pusher-, puller-, or neutral-type flow fields. The pusher-to-puller crossover can be achieved for both outward- and inward-pumping colloids.


We recognize the importance of accurately implementing the no-slip fluid–solid boundary conditions at the colloidal surface without introducing artifacts, in order to generate the desired flow fields around the active colloids. To this end, we introduced an improved approach for enforcing the no-slip boundary condition. The proposed hybrid boundary-condition scheme effectively imposes no-slip at the fluid–solid interface while simultaneously providing an efficient surface thermostat without considering any wall-particles. We have shown that this approach is an improvement over other boundary conditions in providing an accurate no-slip boundary condition. We systematically tested the method and demonstrated its effectiveness through the calculation of several statistical quantities, including verification of the Stokes–Einstein relation.

The present study focuses exclusively on a simple spherical colloid. However, the framework can be readily extended to investigate flow fields of non-spherical colloids as well, with one or more patches. Furthermore, our current analysis is limited to the flow profiles generated by a single colloid suspended in a bulk fluid, but the approach can naturally be generalized to systems containing multiple interacting colloids~\cite{sahoo2021role,thakur2011dynamics,jaiswal2024diffusiophoretic}. The success of the Hybrid boundary conditions also opens the way for theoretical investigation of wet active matter in complex confining geometries, including confining spheres of decreasing radii. Since hydrodynamics is correctly resolved also in the near field,  the consequence of different colloids,  each creating its own distinct hydrodynamic fields, can be investigated, especially for systems of intermediate density, i.e., as long as lubrication forces do not dominate.  The resulting modifications in colloidal dynamics, together with the associated flow fields, are being explored and will be reported in the near future. Moreover, in our study, GPU-enabled computations enabled the calculation of well-averaged flow fields and reliable mean-square displacement calculations.

\section{Acknowledgments}
AC  with DST-SERB (IN) identification No. SQUID-1973-AC-4067, and VC  
acknowledges funding by DST-SERB (IN) project CRG/2021/007824. 
AC also acknowledges discussions in the meetings organized by ICTS, Bangalore, India, 
and the use of the computing facilities by PARAM-BRAHMA, other than departmental  
funds provided by IISER-Pune.

\section{Author Declarations}
\subsection{Conflict of Interest}
The authors have no conflicts to disclose.

\subsection{Author Contributions}
All the calculations and model development presented in the manuscript were performed by OV. The GPU parallelization of the code was implemented by SSRTN with suitable training provided by HG and MM. The problem statement was designed by OV, AC, and RC, and the project was supervised by RC, VKC, and AC. OV, RC, and AC wrote the manuscript, and VKC contributed to the writing of the manuscript. 

\section{Data Availability}
The data that support the findings of this article are openly available~\cite{github_data}. The code can be shared for inspection on a reasonable request.



\appendix

\section{Expressions for viscosity in angular momentum conserving SRD dynamics}
\label{appendix_a}

In this appendix, we reproduce  the analytical expression for the viscosity of an SRD fluid in the angular-momentum-conserving case from~\cite{noguchi2008transport}. The total viscosity, $\eta$, can be decomposed into the sum of kinetic and collisional contributions, $\eta = \eta_{\mathrm{kin}} + \eta_{\mathrm{col}}$. The origin of the kinetic component of viscosity ($\eta_{\mathrm{kin}}$) is in the momentum transport during the streaming step, whereas the collisional viscosity ($\eta_{\mathrm{col}}$) originates from momentum exchange during the collision step. The kinematic
viscosty $ \eta/\rho$ is the diffusion constant for momentum diffusion through the fluid. 

\begin{equation}
\label{eta_kin+a}
\eta_{kin} = \frac{\gamma k_bT \Delta t}{a^3} \left[ \frac{1}{c_m} - \frac{1}{2} \right],
\end{equation}
\begin{multline}
\label{eta_col+a}
\eta_{col} = \frac{m(1-\cos{\alpha})}{36a\Delta t} \times \\  \\ \left( \gamma -\frac{7}{5}+  e^{- \gamma} \left( \frac{7}{5} + \frac{2\gamma}{5}-\frac{3 \gamma^{2}}{10} \right) \right)
\end{multline} \\

with,

\begin{equation}
\begin{aligned}
    c_m = & \frac{2}{5} \left[2-\cos{\alpha}-\cos{2\alpha}\right] \left[1-e^{-\gamma}(1+\gamma)\right] + \\ & \left[ 2(1-\cos{\alpha}) - \frac{11}{5}(2-\cos{\alpha}-\cos{2\alpha}) \right]  \\ & \left[ \frac{1-e^{-\gamma}(1+\gamma+\frac{\gamma^2}{2})}{2 \gamma} \right]
\end{aligned}
\end{equation}

\section{Four random numbers in Eq. 10}
\label{appendix_b}

\begin{figure}[t]
    \includegraphics[width=0.80\linewidth]{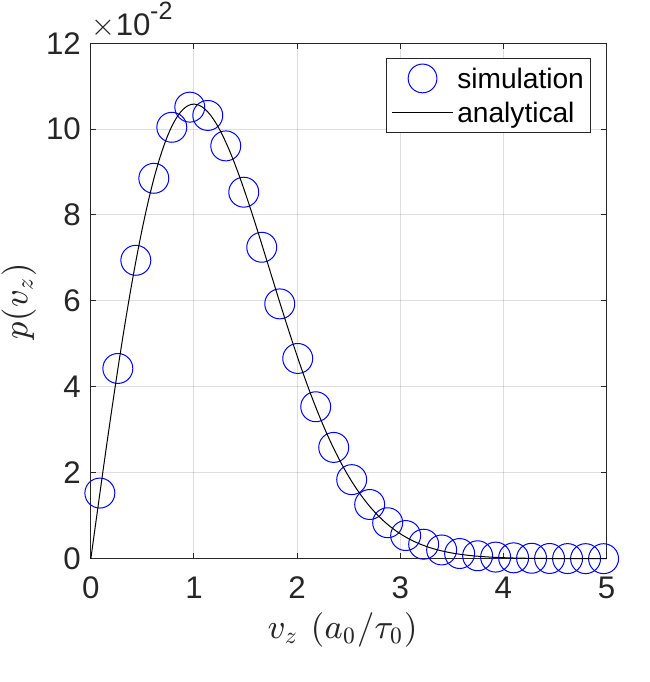}
    \caption{\justifying Probability distribution of the velocity of particles crossing the plane $z = 30 \ a_0$. The figure shows the distribution of the normal ($z$) component of the velocity obtained from simulations using the SRD+a collision scheme in a system of size $30a \times 30a \times 30a$. The measured distribution is consistent with that of a two-dimensional Maxwell–Boltzmann distribution. The corresponding analytical prediction for the two-dimensional Maxwell–Boltzmann distribution is also included for comparison.}
    \label{fig:mb_for_wall}
\end{figure}   

For the stochastic boundary condition described in Sec.~\ref{Boundary condition}, four independent random numbers are employed to generate the post-collision velocity, $\mathbf{v'_i}$. Each of these random numbers is drawn from a Gaussian distribution with zero mean and variance $\sqrt{k_B T / m}$. A straightforward inspection of Eq.(\ref{eq:stochastic_update}) suggests that the resulting average kinetic energy associated with $\mathbf{v'_i}$ is $2 k_B T$. In contrast, the equilibrium average kinetic energy of particles in the bulk fluid is $\tfrac{3}{2} k_B T$. At first glance, this discrepancy might imply a systematic increase in the bulk kinetic energy over time. However, our simulations demonstrate that this is not the case: despite the use of Eq.(\ref{eq:stochastic_update}) at the boundary, the bulk fluid consistently maintains the expected equilibrium value of $\tfrac{3}{2} k_B T$.

This apparent inconsistency can be understood through the following observation. We performed MPC simulations using the SRD+a collision scheme with periodic boundary conditions imposed in all three spatial directions. During the simulation, we recorded the velocities of particles crossing a fixed plane with normal in $z$ direction and analyzed their distributions. The tangential velocity components (in $x$ and $y$ direction) were found to follow Gaussian distributions with zero mean and variance $\sqrt{k_B T / m}$, consistent with equilibrium expectations. In contrast, the normal component (in $z$ direction) exhibited a distribution corresponding to that of a two-dimensional Maxwell–Boltzmann gas at $k_B T = 1~ k_BT_0$ (refer Fig.~\ref{fig:mb_for_wall}). This behavior was consistently observed across different simulation schemes, including SRD-a and molecular dynamics, and remained robust under variation of simulation parameters.

These results indicate that a boundary effectively acts as a statistical filter on particle velocities. For a bulk system characterized by an average kinetic energy of $\tfrac{3}{2} k_B T$, particles crossing a boundary—whether planar or curved—retain Gaussian-distributed tangential velocity components with zero mean and variance $\sqrt{k_B T / m}$. However, their normal velocity component follows the distribution associated with a two-dimensional Maxwell–Boltzmann ensemble. Notably, this same distribution governs particle velocities during stochastic reflections at the wall in our simulations.

During our simulations, when an SRD particle collides with the wall it will follow this very speed distribution. Consequently, to ensure physical consistency, the post-collision velocity $\mathbf{v'_i}$ must be generated in accordance with this distribution. This provides a justification for the use  of Eq.(\ref{eq:stochastic_update}), especially to  generate the normal component of the velocity.


\bibliographystyle{unsrt}
\bibliography{ref.bib}

@article{gompper2009multi,
  title={Multi-particle collision dynamics: A particle-based mesoscale simulation approach to the hydrodynamics of complex fluids},
  author={Gompper, G and Ihle, T and Kroll, DM and Winkler, RG},
  journal={Advanced computer simulation approaches for soft matter sciences III},
  pages={1--87},
  year={2009},
  publisher={Springer}
}

@article{lauga2016bacterial,
  title={Bacterial hydrodynamics},
  author={Lauga, Eric},
  journal={Annual Review of Fluid Mechanics},
  volume={48},
  number={1},
  pages={105--130},
  year={2016},
  publisher={Annual Reviews}
}

@article{ganguly2025hydrodynamics,
  title={Hydrodynamics of squirmers: A review on Stokesian principles, analytical frameworks, and recent advances},
  author={Ganguly, Sourav and others},
  journal={Physics of Fluids},
  volume={37},
  number={12},
  year={2025},
  publisher={AIP Publishing}
}

@article{ganguly2026hydrodynamics,
  title={Hydrodynamics of biomicroswimmers: A Comprehensive review of theoretical foundations, locomotion strategies, and emerging frontiers},
  author={Ganguly, Sourav and Raj, Kiran},
  journal={European Journal of Mechanics-B/Fluids},
  pages={204529},
  year={2026},
  publisher={Elsevier}
}

@article{gompper20252025,
  title={The 2025 motile active matter roadmap},
  author={Gompper, Gerhard and Stone, Howard A and Kurzthaler, Christina and Saintillan, David and Peruani, Fernado and Fedosov, Dmitry A and Auth, Thorsten and Cottin-Bizonne, Cecile and Ybert, Christophe and Cl{\'e}ment, Eric and others},
  journal={Journal of Physics: Condensed Matter},
  volume={37},
  number={14},
  pages={143501},
  year={2025},
  publisher={IOP Publishing}
}

@article{brady2021phoretic,
  title={Phoretic motion in active matter},
  author={Brady, John F},
  journal={Journal of Fluid Mechanics},
  volume={922},
  pages={A10},
  year={2021},
  publisher={Cambridge University Press}
}

@incollection{saintillan2014theory,
  title={Theory of active suspensions},
  author={Saintillan, David and Shelley, Michael J},
  booktitle={Complex fluids in biological systems: Experiment, theory, and computation},
  pages={319--355},
  year={2014},
  publisher={Springer}
}

@article{lighthill1952squirming,
  title={On the squirming motion of nearly spherical deformable bodies through liquids at very small Reynolds numbers},
  author={Lighthill, Michael James},
  journal={Communications on pure and applied mathematics},
  volume={5},
  number={2},
  pages={109--118},
  year={1952},
  publisher={Wiley Online Library}
}

@article{pak2014generalized,
  title={Generalized squirming motion of a sphere},
  author={Pak, On Shun and Lauga, Eric},
  journal={Journal of Engineering Mathematics},
  volume={88},
  number={1},
  pages={1--28},
  year={2014},
  publisher={Springer}
}

@article{goldstein2015green,
  title={Green algae as model organisms for biological fluid dynamics},
  author={Goldstein, Raymond E},
  journal={Annual review of fluid mechanics},
  volume={47},
  number={1},
  pages={343--375},
  year={2015},
  publisher={Annual Reviews}
}

@article{popescu2025hydrodynamic,
  title={Hydrodynamic Stokes flow induced by a chemically active patch imprinted on a planar wall},
  author={Popescu, Mihail N and Nicola, Bogdan A and Uspal, William E and Dom{\'\i}nguez, Alvaro and G{\'a}sp{\'a}r, Szilveszter},
  journal={Journal of Colloid and Interface Science},
  volume={690},
  pages={137296},
  year={2025},
  publisher={Elsevier}
}

@article{anjali2018shape,
  title={Shape-anisotropic colloids at interfaces},
  author={Anjali, Thriveni G and Basavaraj, Madivala G},
  journal={Langmuir},
  volume={35},
  number={1},
  pages={3--20},
  year={2018},
  publisher={ACS Publications}
}

@article{noguchi2008transport,
  title={Transport coefficients of off-lattice mesoscale-hydrodynamics simulation techniques},
  author={Noguchi, Hiroshi and Gompper, Gerhard},
  journal={Physical Review E—Statistical, Nonlinear, and Soft Matter Physics},
  volume={78},
  number={1},
  pages={016706},
  year={2008},
  publisher={APS}
}

@article{gotze2007relevance,
  title={Relevance of angular momentum conservation in mesoscale hydrodynamics simulations},
  author={G{\"o}tze, Ingo O and Noguchi, Hiroshi and Gompper, Gerhard},
  journal={Physical Review E—Statistical, Nonlinear, and Soft Matter Physics},
  volume={76},
  number={4},
  pages={046705},
  year={2007},
  publisher={APS}
}

@article{tn2023gpu,
  title={GPU-based Multiscale Simulation to Model Active Matter Hydrodynamics in Fluid Medium},
  author={TN, SUHAL SIVA RATAN},
  year={2023}
}

@article{bolintineanu2012no,
  title={No-slip boundary conditions and forced flow in multiparticle collision dynamics},
  author={Bolintineanu, Dan S and Lechman, Jeremy B and Plimpton, Steven J and Grest, Gary S},
  journal={Physical Review E—Statistical, Nonlinear, and Soft Matter Physics},
  volume={86},
  number={6},
  pages={066703},
  year={2012},
  publisher={APS}
}

@article{whitmer2010fluid,
  title={Fluid--solid boundary conditions for multiparticle collision dynamics},
  author={Whitmer, Jonathan K and Luijten, Erik},
  journal={Journal of Physics: Condensed Matter},
  volume={22},
  number={10},
  pages={104106},
  year={2010}
}

@article{zottl2018simulating,
  title={Simulating squirmers with multiparticle collision dynamics},
  author={Z{\"o}ttl, Andreas and Stark, Holger},
  journal={The European Physical Journal E},
  volume={41},
  number={5},
  pages={61},
  year={2018},
  publisher={Springer}
}

@article{padding2005stick,
  title={Stick boundary conditions and rotational velocity auto-correlation functions for colloidalparticles in a coarse-grained representation of the solvent},
  author={Padding, JT and Wysocki, A and L{\"o}wen, H and Louis, AA},
  journal={Journal of physics: Condensed matter},
  volume={17},
  number={45},
  pages={S3393},
  year={2005},
  publisher={IOP Publishing}
}

@article{lamura2001multi,
  title={Multi-particle collision dynamics: Flow around a circular and a square cylinder},
  author={Lamura, Antonio and Gompper, Gerhard and Ihle, Thomas and Kroll, DM},
  journal={Europhysics Letters},
  volume={56},
  number={3},
  pages={319},
  year={2001},
  publisher={IOP Publishing}
}

@article{drescher2009dancing,
  title={Dancing volvox: hydrodynamic bound states of swimming algae},
  author={Drescher, Knut and Leptos, Kyriacos C and Tuval, Idan and Ishikawa, Takuji and Pedley, Timothy J and Goldstein, Raymond E},
  journal={Physical review letters},
  volume={102},
  number={16},
  pages={168101},
  year={2009},
  publisher={APS}
}

@article{drescher2010direct,
  title={Direct measurement of the flow field around swimming microorganisms},
  author={Drescher, Knut and Goldstein, Raymond E and Michel, Nicolas and Polin, Marco and Tuval, Idan},
  journal={Physical Review Letters},
  volume={105},
  number={16},
  pages={168101},
  year={2010},
  publisher={APS}
}

@article{drescher2011fluid,
  title={Fluid dynamics and noise in bacterial cell--cell and cell--surface scattering},
  author={Drescher, Knut and Dunkel, J{\"o}rn and Cisneros, Luis H and Ganguly, Sujoy and Goldstein, Raymond E},
  journal={Proceedings of the National Academy of Sciences},
  volume={108},
  number={27},
  pages={10940--10945},
  year={2011},
  publisher={National Academy of Sciences}
}

@article{lushi2014fluid,
  title={Fluid flows created by swimming bacteria drive self-organization in confined suspensions},
  author={Lushi, Enkeleida and Wioland, Hugo and Goldstein, Raymond E},
  journal={Proceedings of the National Academy of Sciences},
  volume={111},
  number={27},
  pages={9733--9738},
  year={2014},
  publisher={National Academy of Sciences}
}

@article{lauga2009hydrodynamics,
  title={The hydrodynamics of swimming microorganisms},
  author={Lauga, Eric and Powers, Thomas R},
  journal={Reports on progress in physics},
  volume={72},
  number={9},
  pages={096601},
  year={2009},
  publisher={IOP Publishing}
}

@article{blake1971spherical,
  title={A spherical envelope approach to ciliary propulsion},
  author={Blake, John R},
  journal={Journal of Fluid Mechanics},
  volume={46},
  number={1},
  pages={199--208},
  year={1971},
  publisher={Cambridge University Press}
}

@article{li2026microrobots,
  title={Microrobots for pulmonary drug delivery},
  author={Li, Zhengxing and Luan, Hao and Fang, Zheng and Ding, Shichao and Kobrin, Robert and Fang, Ronnie H and Zhang, Liangfang and Wang, Joseph},
  journal={Nature Reviews Bioengineering},
  pages={1--17},
  year={2026},
  publisher={Nature Publishing Group UK London}
}

@article{wang2026advanced,
  title={Advanced biofabrication techniques of muscle cell-powered biohybrid robots},
  author={Wang, Niyou and Yang, Yipei and Rezaei, Zahra and Hern{\'a}ndez, Mar{\'\i}a Jos{\'e} Veana and Govindaraj, Kannan and Garzon, Carolina Vazquez and Colin, Marina and Rodr{\'\i}guez, Alan de Jesus Alarcon and Blanco, Alvaro Dario Martinez and Velasco, Jose Joaquin and others},
  journal={International Journal of Extreme Manufacturing},
  volume={8},
  number={1},
  pages={012007},
  year={2026},
  publisher={IOP Publishing}
}

@article{palagi2018bioinspired,
  title={Bioinspired microrobots},
  author={Palagi, Stefano and Fischer, Peer},
  journal={Nature Reviews Materials},
  volume={3},
  number={6},
  pages={113--124},
  year={2018},
  publisher={Nature Publishing Group UK London}
}

@article{pedley2016spherical,
  title={Spherical squirmers: models for swimming micro-organisms},
  author={Pedley, Timothy J},
  journal={IMA Journal of Applied Mathematics},
  volume={81},
  number={3},
  pages={488--521},
  year={2016},
  publisher={Oxford University Press}
}

@article{inoue2002development,
  title={Development of a simulation model for solid objects suspended in a fluctuating fluid},
  author={Inoue, Yasuhiro and Chen, Yu and Ohashi, Hirotada},
  journal={Journal of statistical physics},
  volume={107},
  number={1},
  pages={85--100},
  year={2002},
  publisher={Springer}
}

@article{noguchi2007particle,
  title={Particle-based mesoscale hydrodynamic techniques},
  author={Noguchi, Hiroshi and Kikuchi, Norio and Gompper, Gerhard},
  journal={EPL (Europhysics Letters)},
  volume={78},
  number={1},
  pages={10005},
  year={2007}
}

@article{hecht2005simulation,
  title={Simulation of claylike colloids},
  author={Hecht, Martin and Harting, Jens and Ihle, Thomas and Herrmann, Hans J},
  journal={Physical Review E—Statistical, Nonlinear, and Soft Matter Physics},
  volume={72},
  number={1},
  pages={011408},
  year={2005},
  publisher={APS}
}

@article{malevanets1999mesoscopic,
  title={Mesoscopic model for solvent dynamics},
  author={Malevanets, Anatoly and Kapral, Raymond},
  journal={The Journal of chemical physics},
  volume={110},
  number={17},
  pages={8605--8613},
  year={1999},
  publisher={American Institute of Physics}
}

@book{allen2017computer,
  title={Computer simulation of liquids},
  author={Allen, Michael P and Tildesley, Dominic J},
  year={2017},
  publisher={Oxford university press}
}

@article{goldstein_2024,
  title={Feeders and expellers, two types of animalcules with outboard cilia, have distinct surface interactions},
  author={Prakash, Praneet and Vona, Marco and Goldstein, Raymond E},
  journal={Physical Review Fluids},
  volume={9},
  number={10},
  pages={103101},
  year={2024},
  publisher={APS}
}

@article{wang2019active,
  title={Active patchy colloids with shape-tunable dynamics},
  author={Wang, Zuochen and Wang, Zhisheng and Li, Jiahui and Cheung, Simon Tsz Hang and Tian, Changhao and Kim, Shin-Hyun and Yi, Gi-Ra and Ducrot, Etienne and Wang, Yufeng},
  journal={Journal of the American Chemical Society},
  volume={141},
  number={37},
  pages={14853--14863},
  year={2019},
  publisher={ACS Publications}
}

@article{xiao2025ionic,
  title={Ionic diffusiophoresis of active colloids via galvanic exchange reactions},
  author={Xiao, Zuyao and Simmchen, Juliane and Pagonabarraga, Ignacio and De Corato, Marco},
  journal={Nano Letters},
  volume={25},
  number={19},
  pages={7975--7980},
  year={2025},
  publisher={ACS Publications}
}

@article{rohde2025programmable,
  title={Programmable Hydrodynamics of Active Particles},
  author={Rohde, Lisa and Anchutkin, Gordei and Holubec, Viktor and Cichos, Frank},
  journal={arXiv preprint arXiv:2512.20752},
  year={2025}
}

@article{lauga2006microfluidics,
  title={Microfluidics: The no-slip boundary condition},
  author={Lauga, Eric and Brenner, Michael P and Stone, Howard A},
  journal={Perspective},
  volume={17},
  pages={1},
  year={2006}
}

@article{sahoo2021role,
  title={Role of viscoelasticity on the dynamics and aggregation of chemically active sphere-dimers},
  author={Sahoo, Soudamini and Singh, Sunil Pratap and Thakur, Snigdha},
  journal={Physics of Fluids},
  volume={33},
  number={1},
  year={2021},
  publisher={AIP Publishing}
}

@article{thakur2011dynamics,
  title={Dynamics of self-propelled nanomotors in chemically active media},
  author={Thakur, Snigdha and Kapral, Raymond},
  journal={The Journal of chemical physics},
  volume={135},
  number={2},
  year={2011},
  publisher={AIP Publishing}
}

@article{jaiswal2024diffusiophoretic,
  title={Diffusiophoretic Brownian dynamics: characterization of hydrodynamic effects for an active chemoattractive polymer},
  author={Jaiswal, Surabhi and Ripoll, Marisol and Thakur, Snigdha},
  journal={Macromolecules},
  volume={57},
  number={15},
  pages={6968--6978},
  year={2024},
  publisher={ACS Publications}
}

@article{scagliarini2020unravelling,
  title={Unravelling the role of phoretic and hydrodynamic interactions in active colloidal suspensions},
  author={Scagliarini, Andrea and Pagonabarraga, Ignacio},
  journal={Soft Matter},
  volume={16},
  number={38},
  pages={8893--8903},
  year={2020},
  publisher={Royal Society of Chemistry}
}

@article{de2020self,
  title={Self-propulsion of active colloids via ion release: Theory and experiments},
  author={De Corato, Marco and Arqu{\'e}, Xavier and Pati{\~n}o, Tania and Arroyo, Marino and S{\'a}nchez, Samuel and Pagonabarraga, Ignacio},
  journal={Physical review letters},
  volume={124},
  number={10},
  pages={108001},
  year={2020},
  publisher={APS}
}

@article{lobaskin2004new,
  title={A new model for simulating colloidal dynamics},
  author={Lobaskin, Vladimir and D{\"u}nweg, Burkhard},
  journal={New Journal of Physics},
  volume={6},
  number={1},
  pages={54--54},
  year={2004}
}

@article{chatterji2005combining,
  title={Combining molecular dynamics with Lattice Boltzmann: A hybrid method for the simulation of (charged) colloidal systems},
  author={Chatterji, Apratim and Horbach, J{\"u}rgen},
  journal={The Journal of chemical physics},
  volume={122},
  number={18},
  year={2005},
  publisher={AIP Publishing}
}

@article{palagi2016structured,
  title={Structured light enables biomimetic swimming and versatile locomotion of photoresponsive soft microrobots},
  author={Palagi, Stefano and Mark, Andrew G and Reigh, Shang Yik and Melde, Kai and Qiu, Tian and Zeng, Hao and Parmeggiani, Camilla and Martella, Daniele and Sanchez-Castillo, Alberto and Kapernaum, Nadia and others},
  journal={Nature materials},
  volume={15},
  number={6},
  pages={647--653},
  year={2016},
  publisher={Nature Publishing Group UK London}
}

@article{liebchen2018synthetic,
  title={Synthetic chemotaxis and collective behavior in active matter},
  author={Liebchen, Benno and Lowen, Hartmut},
  journal={Accounts of chemical research},
  volume={51},
  number={12},
  pages={2982--2990},
  year={2018},
  publisher={ACS Publications}
}

@article{ju2025technology,
  title={Technology roadmap of micro/nanorobots},
  author={Ju, Xiaohui and Chen, Chuanrui and Oral, Cagatay M and Sevim, Semih and Golestanian, Ramin and Sun, Mengmeng and Bouzari, Negin and Lin, Xiankun and Urso, Mario and Nam, Jong Seok and others},
  journal={ACS nano},
  volume={19},
  number={27},
  pages={24174--24334},
  year={2025},
  publisher={ACS Publications}
}

@article{ma2015enzyme,
  title={Enzyme-powered hollow mesoporous Janus nanomotors},
  author={Ma, Xing and Jannasch, Anita and Albrecht, Urban-Raphael and Hahn, Kersten and Miguel-L{\'o}pez, Albert and Schaffer, Erik and S{\'a}nchez, Samuel},
  journal={Nano letters},
  volume={15},
  number={10},
  pages={7043--7050},
  year={2015},
  publisher={ACS Publications}
}

@article{tang2020enzyme,
  title={Enzyme-powered Janus platelet cell robots for active and targeted drug delivery},
  author={Tang, Songsong and Zhang, Fangyu and Gong, Hua and Wei, Fanan and Zhuang, Jia and Karshalev, Emil and Esteban-Fern{\'a}ndez de {\'A}vila, Berta and Huang, Chuying and Zhou, Zhidong and Li, Zhengxing and others},
  journal={Science Robotics},
  volume={5},
  number={43},
  pages={eaba6137},
  year={2020},
  publisher={American Association for the Advancement of Science}
}

@article{tang2024bacterial,
  title={Bacterial outer membrane vesicle nanorobot},
  author={Tang, Songsong and Tang, Daitian and Zhou, Houhong and Li, Yangyang and Zhou, Dewang and Peng, Xiqi and Ren, Chunyu and Su, Yilin and Zhang, Shaohua and Zheng, Haoxiang and others},
  journal={Proceedings of the National Academy of Sciences},
  volume={121},
  number={30},
  pages={e2403460121},
  year={2024},
  publisher={National Academy of Sciences}
}

@misc{Supplemental,
title = {Fluid flow profiles for other values of $\theta$ is available in suuplementary information at [url].}
}

@misc{github_data,
howpublished = {\url{https://github.com/captn-nem0/paper1_data.git}},
author = {Vandra, Om and Chatterji, Apratim},
title = {{Hydrodynamics flow fields around an active patchy colloid, GitHub Repository}},
year = {2026},
}

\end{document}